\documentclass[11pt,a4paper]{article}
\usepackage{iftex}
\ifXeTeX
  \usepackage{fontspec}
\else
  \usepackage[T1]{fontenc}
  \usepackage[utf8]{inputenc}
  \usepackage{lmodern}
\fi
\usepackage[margin=2.5cm]{geometry}
\usepackage{booktabs,tabularx,array}
\usepackage{microtype}
\usepackage[hidelinks]{hyperref}
\usepackage{url}
\usepackage{tikz}
\usetikzlibrary{positioning,arrows.meta,decorations.pathreplacing,patterns}
\usepackage{pgfplots}
\pgfplotsset{compat=1.18}
\definecolor{err}{HTML}{EB6834}
\definecolor{cor}{HTML}{2A78D6}
\definecolor{nei}{HTML}{C9C8C2}
\definecolor{ink2}{HTML}{52514E}

\newcolumntype{Y}{>{\raggedright\arraybackslash}X}
\newcolumntype{P}[1]{>{\raggedright\arraybackslash}p{#1}}
\newcolumntype{R}[1]{>{\raggedleft\arraybackslash}p{#1}}
\newcommand{\sect}[1]{\section*{#1}\addcontentsline{toc}{section}{#1}}
\newcommand{\subsect}[1]{\subsection*{#1}\addcontentsline{toc}{subsection}{#1}}
\newcommand{\fp}[1]{{\scriptsize\texttt{#1}}}
\makeatletter
\newcommand{\needlines}[1]{\par\penalty-100\begingroup\dimen@=#1\baselineskip\advance\dimen@\pagetotal
  \ifdim\dimen@>\pagegoal\ifdim\pagetotal>\z@\vfil\penalty-10000\fi\fi\endgroup}
\makeatother

\title{Sealed bets on emerging research topics:\\[0.3em]
\large a protocol for judging early detection, applied to a detector and to eight research funders}
\author{Sébastien Gracia\\
\small Loxyn SAS, Lyon, France\\
\small ORCID \href{https://orcid.org/0009-0001-7885-2655}{0009-0001-7885-2655} \quad \texttt{sebastien@loxyn.ai}}
\date{Preprint \textperiodcentered{} 24 October 2026}

\begin{document}
\maketitle

\begin{abstract}
Tools that claim to detect emerging research topics are rarely judged on their own lists. Our main contribution is a protocol that makes such judgement possible. A bet freezes four things before its verdict can be known: the date of the photograph, the population of candidates, the verdict rule, and the comparison with chance drawn among the same candidates. The list has a fixed length, is read on two axes (champions and failures) against a computed ceiling, and is sealed by cryptographic fingerprint; sealing proves that a list existed at a date, not that it is right. A topic is a champion if its yearly volume of records triples within three years and reaches 100, a failure if it falls by 30\%. We apply the protocol to one detector on PubMed/MEDLINE, in three separate blocks. On a retrospective benchmark (2017--2022), chance among the candidates yields 26.7 champions and 15.1 failures out of 120 slots; the detector yields 85 and 3, and under nine pre-registered variants of the verdict rule its champions stay between 2.5 and 5.0 times chance, with no more failures. The detector was developed with these verdicts known, and its specification is not published. Prospective validation rests on six sealed bets, judged in early 2029 and early 2030; only bets 5 and 6 test exactly the instrument of the benchmark. A secondary, exploratory study applies the same judge to eight research funders: under exact name matching, which may disadvantage grant texts, the detector's lists are more concentrated in future champions than the twenty topics named in the most grants of each funder (70.8\% against 20.8\% to 35.8\%, chance 22.3\%). We document data defects, negative results and the limits of each comparison.

\end{abstract}

\sect{1\texorpdfstring{\quad}{ } Introduction}
\subsect{1.1\texorpdfstring{\quad}{ } A dated birth}
In 2011, hepatitis C was treated by a year of interferon-based therapy that cured about one patient in two. In April and May 2013 the pivotal trials of sofosbuvir were published, with cure rates above 90\% after twelve weeks of tablets. On 6 December 2013 the US Food and Drug Administration approved the drug.

Our detector, run retrospectively on the literature available at each date, dates the rise of this topic to \textbf{July 2011}. At that date the molecule did not yet carry its name. The eight records in PubMed all used its code name, PSI-7977. Four months later, on 21 November 2011, Gilead announced the purchase of Pharmasset, the company that developed it.

The reference first, then the result. The approval came twenty-nine months after the dating. The lead depends on how the dating is computed: across four reference choices recorded before this paper, it ranges from seven to twenty-nine months. All four precede the approval.

This case was chosen because it is famous. It was analysed after the fact. \textbf{A case chosen in this way demonstrates nothing.} It only makes concrete what ``dating a birth'' means. The question this paper asks is different and harder: when a tool says that a topic is emerging, how do we know whether it was right, and how often?

\subsect{1.2\texorpdfstring{\quad}{ } The problem: lists without verdicts}
Many organisations publish lists of emerging topics. Commercial analytics platforms, public foresight programmes and academic methods all produce them. In a survey of public sources made in September 2026 (\S{}2), we found no commercial provider and no public programme that publishes a verdict on its own past lists. Academic work usually reports retrospective splits, where the method is tuned and tested on a known past. Only a few lines of work score their predictions against an explicit external reference, and we found only one that froze its criteria at a date and then judged them out of sample.

The difficulty is structural. A verdict on an emerging topic takes years. In the meantime, it is easy to remember the hits and forget the rest. And without a stated base rate, a hit means little: if most topics grew, finding growing topics would be trivial.

\subsect{1.3\texorpdfstring{\quad}{ } Contributions}
This paper makes one main contribution and applies it.

\begin{enumerate}
\item \textbf{The protocol is the main contribution} (\S{}3). A bet freezes four things before its verdict can be known: \textbf{the date} of the photograph; \textbf{the population of candidates} from which the list is drawn; \textbf{the verdict rule}; and \textbf{the comparison with chance drawn among the same candidates}, with a computed ceiling, on two axes (champions and failures), in declared units. The list has a fixed length and is sealed by fingerprint. Three outcomes are written in advance. The protocol does not depend on the detector, and it stays defensible if a bet fails: a failed bet, judged by a rule written in advance, is a result the protocol is built to produce.
\item \textbf{An experimental application of the protocol: one detector} (\S{}5--\S{}11), on PubMed/MEDLINE, in three blocks kept apart (\S{}1.4): exploration, historical validation on a retrospective benchmark (2017--2022), and prospective validation by sealed bets.
\item \textbf{The long-term verification: six sealed bets} (\S{}11), whose fingerprints are timestamped in the Bitcoin blockchain through OpenTimestamps. Five will be judged on the 2028 volumes, read once the indexing of 2028 is complete (early 2029); one on the 2029 volumes, in early 2030.
\item \textbf{A secondary, exploratory study: research funders} (\S{}12). The same judge, applied to the grant portfolios of eight funders in Europe and the United States, through name matching.
\end{enumerate}

We also report the data defects we found (\S{}4) and what did not work (\S{}6).

The detector itself is not the contribution. Its specification is not published; \S{}7.2 explains why, and what can nevertheless be judged from outside.

\subsect{1.4\texorpdfstring{\quad}{ } Three blocks, kept apart}
The results of this paper are of three kinds, and they must not be confused.

\begin{itemize}
\item \textbf{Exploration} (\S{}5--\S{}6) reports what was tried, including what failed. It supports no claim about the detector; it tells the reader how many paths were tried.
\item \textbf{Historical validation} (\S{}7--\S{}10) judges the detector on a benchmark of six past vintages, 2017--2022. It is retrospective evidence, developed with the verdicts known (Part B). It cannot, by itself, prove anything about the future.
\item \textbf{Prospective validation} (\S{}11) consists of sealed bets whose outcome nobody knows yet. Only this block does not ask to be believed, and any claim about future performance rests on it.
\end{itemize}

\begin{quote}
\textbf{Which bets test which instrument.} Bets 1 to 4 \textbf{do not test exactly} the instrument that produced the 85/3 of the historical benchmark: each differs from it in parts of the specification. \textbf{Bet 5 is the first bet that tests that exact instrument.} \textbf{Bet 6 tests it on up-to-date data}: the photograph of 30 June 2026, read on the baseline plus the updates to 6 October 2026.
\end{quote}

The register of corrections (Annex A) is filed under the same three blocks.

\sect{2\texorpdfstring{\quad}{ } Related work}
\subsect{2.1\texorpdfstring{\quad}{ } Literature-based discovery}
The idea that a discovery may already lie in the literature, scattered among works that ignore one another, goes back to Swanson [1]. Two sets of articles establish that A affects B and that B affects C; nobody has written that A may act on C. This line of work, literature-based discovery, has worked on MEDLINE for decades: Swanson and Smalheiser with Arrowsmith [2], Wren and colleagues with co-occurrence over the whole of MEDLINE against a random null model [3], Smalheiser and Bonifield with observed-over-expected ratios over millions of descriptor pairs [4]. Moreau [5] discusses how this field evaluates itself. Our question differs. We do not look for a link nobody has made. We look for a topic whose growth nobody has yet noticed.

\subsect{2.2\texorpdfstring{\quad}{ } Measures of novelty and breakthrough}
Three families dominate the quantitative study of novelty. \textbf{Atypical combination} [6] compares the observed frequency of co-cited journal pairs with a randomised citation network; the finding is that a minority intrusion of unusual combinations into otherwise conventional work raises the rate of high-impact papers. Novelty as \textbf{first combination} is measured by Lee, Walsh and Wang [7] and by Wang, Veugelers and Stephan [8]. The \textbf{disruption index} [9] distinguishes work that consolidates from work that destabilises, through the local structure of the citation graph; it was built and validated on patents and later extended to articles [10, 11]. The claim that science is becoming less disruptive [11] is debated: Petersen, Arroyave and Pammolli [12] argue that the index is biased by citation inflation, and an exchange published in \emph{Nature} in August 2026 [13, 14] attributes part of the measured decline to dataset artefacts. We do not take sides in this debate.

\textbf{Burst detection} [15] models a stream of documents as a probabilistic automaton in which a burst is a state transition; it has been applied to bibliographic corpora [16, 17]. These methods describe a burst while it happens. We found no prospective validation of burst detection on PubMed.

\subsect{2.3\texorpdfstring{\quad}{ } Forecasting emerging topics}
Small, Boyack and Klavans [18] identify emerging topics and validate nominated topics by the prizes they later received, without reporting precision or recall. Rotolo, Hicks and Martin [19] give the five-attribute definition of an emerging technology that the field now uses. The line of work by Klavans and Boyack is the only one we found that scores forecasts against an external threshold: in a 2020 study [20], over about 100,000 Scopus communities and a three-year horizon, the threshold inherited from the IARPA FUSE programme (a critical success index of 25\%) was exceeded in six domains out of nine. That paper published ten communities forecast to grow exceptionally after 2018 and announced that they would be scored once data through 2021 were available; we did not find that scoring. A later study extended the horizon to four years [21].

\textbf{AUGUR} [22] forecasts the emergence of new research topics, before they are named, from the increasing pace of collaboration between ancestor topics in the Computer Science Ontology [23]. Clusters found in year \emph{t} are matched with topics that debut (reach five publications) in the following two years; on a gold standard of 1,408 debutant topics (2000--2011), it outperformed four alternatives, with precision up to 0.86 and recall up to 0.76 in early years and lower values in recent years, which the authors attribute to topics that had not yet debuted in the gold standard. The same group studied the research dynamics that precede the birth of a topic [39] and forecast the entry of new concepts into an ontology [40], which is close to our use of descriptor creation as a clock (\S{}11.5). Khanbayov and Kurban [24] froze four criteria on 2017--2021, evaluated them out of sample on 2023--2025 (precision 27\%, recall 63\%, base rate 13.5\%), and published five topics forecast for 2026--2028. This is the closest case we found to a sealed prediction later judged. The work of Krenn and colleagues provides open benchmarks for concept-link prediction [25, 26]; recent benchmarks such as PreScience [27] and ForeSci [28] use strict temporal cut-offs. These are retrospective splits, not sealed predictions. Maillart and colleagues [29] report high ROC-AUC on OpenAlex without a language model. Ofer and Linial [30] forecast trends on PubMed and patents five years ahead. Lu, Yang and Wang [31] use MeSH descriptors added in 2001--2010 as ground truth, with a counterfactual sample of topics that did not take off. Li and colleagues [32] have to construct their own metric for lack of ground truth, which echoes the conclusion of Doos and colleagues [33] that ``the retrospective accuracy of the studies remains to be researched''.

The work closest to ours in corpus comes from a funder. The NIH Office of Portfolio Analysis [41] searched co-citation networks built over PubMed, year by year and with only the information available at each date, for a signature that precedes transformative breakthroughs. On a curated set of 21 breakthroughs associated with major prizes (1984--2014), the signature appears on average more than five years before the papers that announced the discovery, and its kinetics hold across two windows twenty years apart. The evaluation is retrospective, its reference is a set of prize-winning discoveries rather than a population of candidates, and we did not find in it a sealed list of forward predictions.

Tshitoyan and colleagues [34] trained word embeddings on 3.3 million materials-science abstracts and showed that the top fifty candidates of each historical corpus were eight times more likely than a random material to be studied as thermoelectrics within five years, and three times more likely than a random material with a non-zero band gap. They published two baselines. Sourati and Evans [35] predict future discoveries from the literature, accounting for who works on what.

\subsect{2.4\texorpdfstring{\quad}{ } What is missing}
Three things are missing from this landscape. First, \textbf{a verdict on one's own lists}, published after the fact: we found it in academic work only once [24], and nowhere in commercial or public foresight. Second, \textbf{a base rate computed among the same candidates} the method chose from. Publishing a baseline is not new [6, 34], and we do not present it as a contribution. What we add is narrower: the baseline is drawn from the population of the bets themselves, it comes with a ceiling, and the cost of choosing after looking at the answer is measured. Third, \textbf{a cryptographic commitment} that prevents rewriting the list once the verdict is known. Timestamping is old [36]; we only apply it.

\sect{3\texorpdfstring{\quad}{ } The evaluation protocol}
\subsect{3.1\texorpdfstring{\quad}{ } A bet}
A \textbf{bet} has three parts, all fixed before the verdict can be known:

\begin{itemize}
\item \textbf{a date}, the photograph. All candidates are frozen on the same day, with the same information. We use 30 June of each year \emph{Y}.
\item \textbf{a list of fixed length}. We use twenty slots per photograph. A detector that could lengthen its list until it contained a success would prove nothing.
\item \textbf{a judge written in advance}. The rule that decides what each listed topic has become.
\end{itemize}

The term \textbf{slot} denotes one topic placed in one year. A topic can occupy slots in several years. A \textbf{vintage} is the set of slots of one photograph.

\subsect{3.2\texorpdfstring{\quad}{ } The judge}
Let $n_y(g)$ be the number of distinct records carrying topic \emph{g} in year \emph{y}. Three years after the photograph:

\begin{center}
\begin{tabular}{@{}lll@{}}
champion & if & $n(Y{+}3) \geq 3\cdot n(Y{-}1)$ \quad and \quad $n(Y{+}3) \geq 100$\\
failure & if & $n(Y{+}3) \leq 0.7\cdot n(Y{-}1)$\\
otherwise & & no verdict
\end{tabular}
\end{center}

The rule was written before any measurement and is the same for every topic, every method and every funder in this paper. It measures \textbf{growth in the literature}. It does not measure scientific truth, clinical value or commercial success. A fashion that is written about a great deal can be a champion. The rule also measures growth, not birth: an old and generic topic whose literature grows can satisfy it. These are limits of the judge, stated here because every number below depends on it.

\needlines{10}
\textbf{What it costs to pass the judge.} Over all the topics tracked in the corpus, the reference is as follows (sealed file of 1 September 2026, Annex B):

\nopagebreak[4]
\begin{center}\small
\begin{tabular}{@{}lr@{}}
\toprule
what a tracked topic becomes in three years &  \\
\midrule
median growth & $\times$1.18 \\
99 topics in 100 stay below & $\times$3.45 \\
topics that triple & 1.7\% \\
\textbf{champions} (triple \emph{and} reach 100 records) & \textbf{1.4\%}, one in 70 \\
\textbf{failures} (fall by 30\% or more) & \textbf{9.4\%}, about one in eleven \\
\bottomrule
\end{tabular}
\end{center}

Tripling sits near the 99th percentile. Failure is about seven times more frequent than the feat. Its share also changes over the period: 3.6\% of tracked topics failed in the 2017 vintage and 15.8\% in the 2022 vintage, while the share of champions halved. Vintages are therefore never pooled without being shown separately.

\subsect{3.3\texorpdfstring{\quad}{ } Chance among the same candidates}
Any detector designates its topics from a population of eligible candidates. A list of twenty must be compared with what twenty topics drawn at random \textbf{from the same eligible population} would have achieved. Comparing with the whole corpus flatters any detector that restricts its candidates.

The difference is large. On our benchmark (\S{}8), twenty topics per year drawn at random from all tracked topics yield \textbf{1.8 champions and 10.3 failures} over 120 slots. Drawn from the detector's own candidates, they yield \textbf{26.7 champions and 15.1 failures}. The first number is the reference for a reader who asks whether the whole system works. The second is the reference for the question that matters here: does the detector do better than a random draw from its own candidates? All results in this paper are judged against the second.

In percentages, chance among the candidates, weighted like the list (twenty slots per vintage), is 22.3\% champions and 12.6\% failures per slot; unweighted, over all candidate-years, it is 20.2\% and 13.0\% (\S{}12.2). About one candidate in eight becomes a failure. Avoiding failures is therefore not trivial either.

\subsect{3.4\texorpdfstring{\quad}{ } The ceiling}
How many champions would a perfect oracle take? Since only twenty slots are available per year, the answer is the sum over vintages of min(20, champions available). On our benchmark the attainable ceiling is \textbf{114 of 120}. A result is easier to read as a distance to this ceiling than as an absolute score.

\subsect{3.5\texorpdfstring{\quad}{ } The null model of the search itself}
A team that tries many methods and reports the best one can manufacture a result. The protocol therefore includes a null model for its own search. The rankings produced by the whole catalogue of methods we tried are fixed. The verdicts are permuted 20,000 times. At each draw, the best method of the catalogue is kept, which is exactly the selection we could be accused of. The test is hardened by playing each statistic in both directions. Under this null, \textbf{chance tops out at 47 champions out of 120 slots}, never beyond in 20,000 draws.

A second lesson came from our own development. On an earlier state of the benchmark, the best combination of methods chosen after seeing its score gave 88 champion slots; the same procedure run blind gave 82. Choosing while looking was worth six slots. We report the blind figure in such cases.

\subsect{3.6\texorpdfstring{\quad}{ } Units}
Two units coexist and must never be compared with each other. A \textbf{slot} is one topic in one year. A \textbf{distinct topic} may occupy several slots. One topic can also be designated by several names; a counting rule ensures that equivalent names of the same topic do not occupy two slots of the same list. That rule is part of the specification (\S{}7). On our benchmark, the 85 champion slots carry \textbf{63 distinct names (59 after merging equivalent names)}.

We learned this the hard way. An earlier count of 88 champion slots turned out to cover 80 distinct objects once equivalent names were recognised, and an internal note once compared a number of names with a number of slots. Both corrections are in the register (Annex A).

\subsect{3.7\texorpdfstring{\quad}{ } Two axes, and below the cut}
A detector can buy champions with failures. Every result in this paper is therefore reported on two axes, champions and failures, against their chance values. A gain announced without its cost means nothing.

A list of twenty also hides what happens below rank twenty. During development, a candidate addition produced the largest gain in champions of the whole campaign. A control that looks below the cut showed that the gain came at the expense of true champions ranked below the top twenty, where their loss cost no slot in the table. The addition was rejected. Any evaluation by top-\emph{k} should include such a control.

\subsect{3.8\texorpdfstring{\quad}{ } Sealing}
A list is sealed in four steps:

\begin{enumerate}
\item the list file is fingerprinted with SHA-256;
\item for bets 5 and 6, each position is also committed separately: a line (rank, topic, data of the slot), a random salt of 16 bytes, and the fingerprint SHA-256(salt | line); the public file of fingerprints has its own root fingerprint, so that one position can be revealed without revealing the others;
\item for bets 5 and 6, the candidate population is fingerprinted too, so that chance can be computed at the verdict on exactly those candidates;
\item a manifest listing these fingerprints, together with internal file names, is timestamped with OpenTimestamps, which anchors its own fingerprint in the Bitcoin blockchain.
\end{enumerate}

Only fingerprints leave the machine; the manifests themselves are not published. The lists stay closed until their verdict. Earlier versions of a list that were withdrawn before any circulation remain anchored, and each withdrawal is declared in a later manifest; nothing is erased. Anyone can check with \texttt{ots verify} that each anchored manifest existed at its date (\S{}11.4). \textbf{Sealing proves that a list existed at a date, not that it is right.} Only the verdict can say that.

\subsect{3.9\texorpdfstring{\quad}{ } Three outcomes, written in advance}
For every sealed bet, the reading is fixed now:

\begin{itemize}
\item the lists hold the benchmark rate: the benchmark was not optimistic;
\item they beat chance without holding the rate: the benchmark was optimistic;
\item they do not beat chance: everything above was a retrospective artefact, and we will publish it as such.
\end{itemize}

The third outcome gives the other two their meaning. A protocol that cannot lose proves nothing.

\subsect{3.10\texorpdfstring{\quad}{ } Two reservations that the protocol does not remove}
\textbf{The photograph does not freeze the database.} Our substrate is a 2026 snapshot. When we reconstruct what an observer saw on 30 June of a past year, we read records as they are today. Some records dated before the photograph were indexed after it, and some were revised. \S{}4.1 measures how much for one recent period. For future bets the remedy is structural: we plan to keep a dated copy of the substrate at each photograph (\S{}13).

\textbf{The photograph overlaps the judge's denominator.} A photograph on 30 June of \emph{Y} sees the first half of year \emph{Y}, and everything before. The denominator of the judge, \emph{n(Y$-$1)}, is therefore fully visible at the time of the bet. This can create an artificial correlation on the failure axis. The point is real and unresolved.

\sect{4\texorpdfstring{\quad}{ } Data and their defects}
\subsect{4.1\texorpdfstring{\quad}{ } The frozen snapshot}
We work on the whole of PubMed/MEDLINE in its annual baseline distribution, \textbf{frozen on 18 August 2026: 1,334 files, 39,994,988 records}, plus the PubTator 3 entity layer [37] (448 million links). The manifest of the 1,334 files, with their sizes and dates, is sealed (fingerprint \texttt{615912ccc2ee341c\dots{}}) and timestamped. The substrate is opened read-only.

The instrument reads \textbf{information drawn from the records}, dated month by month. We do not say which (\S{}7.2). This section reports defects of PubMed and of the annotation layers built on it, at that level of description.

Why freeze? Because PubMed is a living database. On 7 October 2026 we downloaded the update files issued after the baseline (to file \texttt{n1651}, published on 6 October 2026) to prepare the sixth bet. Over these files, 2,518,536 record versions were retained. \textbf{1,301,505 of them replace a record of the baseline; 1,217,031 are new; 2,326 baseline records are deleted.} These revisions change the instrument's output only slightly (below): the publication date it uses moves for 18,474 of about 1.3 million revised records, almost all from 2025 to 2026.

These are small changes. They are not without effect. Re-running the frozen program on the baseline plus updates, the 2025 photograph gives \textbf{the same twenty topics} as the sealed bet 5, but \textbf{eight of the twenty positions change order}, and the candidate population gains two topics. An internal control of the program changes by one champion. The cause is late records and revisions. We declare this in bet 6 (\S{}11). \textbf{Without a frozen snapshot, a dated measurement does not reproduce.}

Completeness also matters. Three months after the end of June 2025, the month was present at 97.9\% of its current volume. On 7 October 2026, June 2026 was judged complete enough for a photograph: 152,290 records, 1.18 times the volume June 2025 had at the same delay.

\subsect{4.2\texorpdfstring{\quad}{ } Defects found, dated and measured}
During the campaign we recorded forty forms of data defects across PubMed, PubTator and the other corpora we used, in an internal atlas dated 9 September 2026. Each entry carries a dated measurement. We give the ones that matter for anyone building time series on PubMed.

\textbf{Homonymy in the entity layer.} An annotation service maps strings to concepts. Short acronyms rarely belong to one domain. We found the problem while building the sofosbuvir case: the string ``SOF'' is normalised to sofosbuvir, an antiviral approved in 2013, but in reproductive biology it has meant \emph{synthetic oviductal fluid}, an embryo culture medium, since the 1970s. The seed of a molecule first described in 2010 contained records decades older than the molecule. \textbf{At the date of the dating, 26 of the 34 records carrying the concept were false.} The dating of \S{}1.1 was checked against this defect (Annex A, A18).

A systematic search over all 448 million links, comparing the share of each spelling in old and recent records, confirmed \textbf{52 homonymies covering about 2.4 million contaminated lines}, and we estimate that some \textbf{800} more remain to be confirmed. The concept COVID-19 is also attached to records published before 2020, through mentions of the SARS epidemic of 2003 and of the word ``pandemic''. The general rule we draw is simple. \textbf{Read the oldest records of any concept built by string matching. An object born in 2010 cannot have an article from decades before.} And an acronym is not ambiguous in the absolute; it is ambiguous at a date.

\textbf{Retro-indexing of the controlled vocabulary.} A MeSH descriptor created in 2019 is attached to articles from 2006; 931 records completed before 2018 carry it. A dating of an emergence by the first records carrying a descriptor can be wrong by years.

\textbf{A new label for an old science.} A descriptor can ``appear'' with more than a thousand records on the day of its creation (1,411 for one case we measured). A vocabulary event is then read as an emergence.

\textbf{Lag between rise and label.} Conversely, the official vocabulary can create a descriptor long after a topic is visible in the literature. On the five positions of our dated register that are already public (\S{}11.5), the dating precedes the creation of the official descriptor by twelve months to 7.1 years.

\textbf{One event, many nodes.} A single real-world event can produce many neighbouring terms, in a controlled vocabulary, in entity annotations and in free text alike. A novelty measure that counts each new term as a birth counts one event several times.

\textbf{A change of indexing regime.} MeSH indexing moved from human indexers to automated indexing during the period, around 2022. A series of MeSH indexing that spans the switch changes source without saying so: a movement in such a series may be in the procedure, not in the world. Any such series must be stratified by regime. We observed the switch; we have not measured its size.

\textbf{No absolute clock, two dates.} Records carry revision bounds rather than a date of fact; a revision date taken as a date of fact can make a record from the 1940s look fifty years younger. Each record also carries a print date and an electronic date, which can differ by months; the snapshot contains records dated 2027 by advance publication.

\textbf{Citations and corrections.} About 48.5 million citation links point outside the snapshot, and two point to identifiers that cannot exist. 1,429 records are discordant between their retraction status and their links.

\textbf{Affiliations and countries.} Country detection by pattern in affiliations is fragile. Our own first version searched for substrings: ``uk'' matched Duke, Fukuoka, Milwaukee and leukemia. \textbf{31.5\% of the records it marked as British were false.} It was corrected on 11 September 2026. The field giving the country of the funder holds a sentinel value for 72.4\% of the recent funded records we examined.

\textbf{Our own conventions.} Two of our derived tables were built with inconsistent conventions (703 entries affected). The defect was measured as having no effect on the lists derived from the whole world corpus. It has not been measured on the retrospective benchmark, where it could touch only the 2021 and 2022 verdicts.

\subsect{4.3\texorpdfstring{\quad}{ } What these defects mean for evaluation}
Two of these defects create false emergences: retro-indexing and vocabulary events. Two create false declines: late indexing and revisions (\S{}4.1). One, homonymy, creates false history. None is visible on an aggregate curve. A protocol for judging emergence must therefore fix its corpus by fingerprint, state its dates of reading, and publish the controls it applied.

\sect{Part A --- Exploration: what was tried}
This block reports the paths we tried around the detector: public methods under the protocol (\S{}5) and what did not work (\S{}6). It supports no claim about the detector. Its purpose is to let the reader judge how many paths were tried before the results of Part B. The funder study (\S{}12), secondary and exploratory, is reported separately.

\sect{5\texorpdfstring{\quad}{ } Public methods under the protocol}
\needlines{12}
\subsect{5.1\texorpdfstring{\quad}{ } Flow methods, on all tracked topics}
We first applied public methods to all tracked topics, with each method's original parameters and no tuning. Twenty slots per year, six vintages, the judge of \S{}3.2:

\nopagebreak[4]
\begin{center}\small
\begin{tabular}{@{}lrr@{}}
\toprule
method, on all tracked topics & champions & failures \\
\midrule
\emph{chance} & \emph{1.8} & \emph{10.3} \\
raw volume of the previous year & 0 & 5 \\
naive growth ratio & 22 & 47 \\
burst detection [15] & 45 & 10 \\
\bottomrule
\end{tabular}
\end{center}

\emph{(120 slots in each line; the counts come from an internal file of 1 September 2026.)}

Three readings. \textbf{Volume alone finds no champion}: its twenty first topics are the largest of the literature, which never triple. \textbf{The best public method is far above chance and unusable}: 45 champions for 10 failures, and 65 of 120 slots that are neither. About one listed topic in three becomes a champion, with as many failures as chance would commit. \textbf{The failure axis is where public methods are weakest}: the naive growth ratio commits more than four times as many failures as chance, and burst detection does no better than chance on that axis.

\subsect{5.2\texorpdfstring{\quad}{ } The catalogue of statistics}
Before building anything of our own, we tried \textbf{262 statistics, 524 candidates and 1,320 pairwise combinations}, across the main families of time-series statistics. None of them, used alone or in pairs, comes close to the result of \S{}8. The null model of \S{}3.5 shows that searching this wide could not, by itself, produce the result of \S{}8.

\subsect{5.3\texorpdfstring{\quad}{ } Structure methods, on the candidates}
The two flagship measures of the science of science apply to articles, not to topics. We averaged them over the articles of each topic. To give them every chance, we measured them \textbf{on the detector's candidates}, where chance is already high, and compared them with chance among those candidates, on both axes.

The disruption index was computed on the citation layer of the substrate (about 398 million links), at the date of the photograph, where we expected it to fail for lack of citing articles, and with all the citations known today, where we expected a retrospective success. \textbf{In either regime it does not beat chance among the candidates.} Atypicality is computed at publication and needs no future citation, so it could in principle predict. \textbf{It does not beat chance among the candidates either.}

The reason is not sparsity. These indices measure \textbf{the nature of a contribution}: does this article break with its antecedents, does it combine unusual sources? Our question concerns \textbf{the trajectory of a topic}. A consolidating article about a topic that explodes is still a consolidating article. Two reservations apply. These are our adaptations that fail, not the original works, whose authors never claimed to date the birth of a topic. And our null law for atypicality is an analytical approximation, coarser than the Monte Carlo resampling of the original [6].

The selection of the candidate population is itself part of the detector, not a preliminary to it: it raises the base rate of champions from 1.4\% of all tracked topics to about 22\% of candidates (\S{}3.3). For that reason we give no count for these measures, nor for any public method run inside this population (\S{}7.2).

\subsect{5.4\texorpdfstring{\quad}{ } What these trials establish}
On all tracked topics, flow methods are the only live channel, and bare they commit many failures. Structure methods are silent, even when given the detector's candidates. Content methods based on text embeddings [34] were not compared, and we do not claim anything about them.

\sect{6\texorpdfstring{\quad}{ } What did not work}
A file that publishes only what works cannot be checked: the reader cannot know how many paths were tried. This section gathers the failures that matter for the protocol.

\subsect{6.1\texorpdfstring{\quad}{ } Closed search spaces}
We declare three search spaces closed, which commits us not to return to them without saying why.

\begin{itemize}
\item \textbf{Time-series statistics.} The catalogue of \S{}5.2: nothing in it comes close to the result of \S{}8.
\item \textbf{Carriers, places, countries.} Several promising signals about who signs, team renewal, concentration, dispersion across journals, spread towards general journals and exit from the country of origin \textbf{turned out to be volume in disguise} once size was neutralised, and some changed sign when stratified. This trap nearly made us publish a gain that was only a size effect.
\item \textbf{External review.} An independent adversarial review proposed four improvements. None improved the result: three were measured and failed, and the fourth could have had no effect, by construction.
\end{itemize}

\subsect{6.2\texorpdfstring{\quad}{ } A dating is not a verdict on the winner}
The dating of \S{}1 says that a topic is rising. It does not say that the topic will win. We tested this on molecules under a protocol written before calculation. On the 125 molecules that could be judged (16 that died, 109 that lived), the early intensity of the signal separates them with an area under the ROC curve of \textbf{0.496} (95\% interval 0.353--0.646): chance. In duels matched by company, period and chemistry, the figure is \textbf{0.413}, worse than chance: the molecule that will die tends to have more early strength than the one that will live.

\subsect{6.3\texorpdfstring{\quad}{ } A theory that did not hold}
We tried to turn the framework into a mathematical theory. Three structural predictions were written before any calculation, with their conditions of refutation. \textbf{Two of the three fell.} We accept the conclusion: what we have is engineering, not a theory.

\subsect{6.4\texorpdfstring{\quad}{ } A validation that read the future}
During development, a revision of the instrument was validated on the benchmark. On 1 September 2026 we found that this validation read the whole series to 2025 for every vintage: a candidate of 2021 was judged with the decline of the COVID-19 literature of 2023--2025 in view. Replayed without the future, the revision fell far below the frozen protocol, almost entirely in the epidemic vintages. The part responsible was replaced. \textbf{Bet 2 had already been sealed with it} (\S{}11). The bet stands as sealed; one does not rewrite a bet.

\subsect{6.5\texorpdfstring{\quad}{ } The detector depends on its material}
The performance reported in this paper depends on the material the detector reads. It was measured on PubMed/MEDLINE only. It is not a property of counting alone, and it should not be assumed to transfer to other material or to another corpus without a new measurement.

\sect{Part B --- Historical validation: the 2017--2022 benchmark}
This block is \textbf{retrospective evidence}, and we say so once, here. The detector was developed over months on this benchmark, with its verdicts known. The verdict rule was written before any measurement, but this cannot be checked from outside. The temporal control of the benchmark is also imperfect, in three ways: the photographs of past years read a snapshot of 2026, in which some records were indexed or revised after the photograph (\S{}3.10, \S{}4.1); some inputs of the program were prepared after the photograph dates; and the photograph overlaps the denominator of the judge (\S{}3.10). Two safeguards bound these doubts without removing them: the null model of our own search (\S{}3.5) and the sensitivity of the judge (\S{}10), under which the advantage holds across nine variants of the verdict rule. The controls of \S{}8.2 add others. No claim about future performance rests on this block: such claims rest on the sealed bets of Part C, and specifically on bets 5 and 6.

\sect{7\texorpdfstring{\quad}{ } The detector evaluated}
\subsect{7.1\texorpdfstring{\quad}{ } What is published}
The following paragraph is the only description of the detector that we publish:

\begin{quote}
The instrument combines \textbf{four organs of the same mathematical species} --- exact counting statistics, without a single learned parameter. \textbf{To these is added a layer of another species}, without a single learned parameter either. \textbf{Twenty constants in all}, fixed before their trials --- nine swept over their entire grid in the full file; one could not be measured, and that is declared; four carry their variants in print; the remaining six are each backed by a measurement. It is in the candidate population --- defined by the instrument --- that chance is worth 26.7 champions and 15.1 failures out of one hundred and twenty slots, and it is against it that everything that follows is judged.
\end{quote}

In addition: the detector runs offline, uses no language model, and is deterministic. On 7 October 2026 we re-ran the program that produced the benchmark of \S{}8, unchanged, on the same inputs, in a new directory. \textbf{The output file was identical byte for byte} to the file sealed on 4 September 2026. It produces lists of a fixed number of places; this paper uses \textbf{twenty places} (\S{}9).

\subsect{7.2\texorpdfstring{\quad}{ } What is not published, and why}
The specification of the detector is \textbf{held by Loxyn SAS and not published}. This covers the material read, the definition of the candidate population, the organs, the layers, the constants and the units. We make this choice for commercial reasons and declare it as such.

The candidate population is part of the detector, not a preliminary to it, and it carries much of the result (\S{}5.3). We therefore publish no component-wise result, and no count for a public method run inside this population (\S{}5.3 reports only that two structure measures do not beat chance there): such results would show how the population is drawn, and would offer a target for rebuilding it.

This creates an obvious difficulty for a scientific paper: a reader cannot rebuild the detector. We address it in four ways.

\begin{itemize}
\item \textbf{The judge's rule is public} (\S{}3.2). Its counts, however, are taken on material and with conventions that are part of the specification. A reader therefore cannot recount our verdicts exactly from PubMed alone. One route remains: a reader can apply the public rule with a volume measure of their own, for example a PubMed query on the name of each listed topic, and declare the gap between that measure and ours.
\item \textbf{The 120 judged slots of the benchmark}, with their verdicts, are available under agreement; their file is fingerprinted (Annex B), so that the copy shown under agreement can be checked against the sealed one.
\item \textbf{The bets do not ask to be believed} (\S{}11). They are sealed before the verdict, and the verdict rule is fixed in advance.
\item \textbf{We tested whether the published results reveal the detector.} Pre-registered blind reverse-engineering tests, with scoring rules fixed before reading, were run on our lists and on this paper.
\end{itemize}

Whether this is acceptable for a scientific claim is a fair question. We have tried to make the claim depend on what can be checked, not on what is withheld.

\sect{8\texorpdfstring{\quad}{ } Results on the 2017--2022 benchmark}
\needlines{12}
\subsect{8.1\texorpdfstring{\quad}{ } The main table}
Six vintages, 2017 to 2022, twenty slots each: \textbf{120 slots}. Outputs sealed by fingerprint (file \texttt{d059c3f0\dots{}}, 4 September 2026). The reference first, then the result:

\nopagebreak[4]
\begin{center}\small
\begin{tabular}{@{}lrr@{}}
\toprule
 & champions & failures \\
\midrule
chance among all tracked topics & 1.8 & 10.3 \\
chance among the detector's candidates & 26.7 & 15.1 \\
attainable ceiling & 114 & --- \\
\textbf{the detector} & \textbf{85} & \textbf{3} \\
\bottomrule
\end{tabular}
\end{center}

\emph{Chance is computed with champions counted as distinct topics after merging equivalent names; the detector's line is counted in slots. The two conventions differ by about one per cent on our benchmarks.}

By vintage, champions/failures: \textbf{12/0 \textperiodcentered{} 13/0 \textperiodcentered{} 18/0 \textperiodcentered{} 15/0 \textperiodcentered{} 14/2 \textperiodcentered{} 13/1.} The 85 champion slots carry \textbf{63 distinct names (59 after merging equivalent names)}.

Read against chance among the candidates, the detector's champion rate is 70.8\% of slots, against 22.3\% for chance: a ratio of 3.2. Its failure rate is 2.5\%, against 12.6\%. From all tracked topics to the detector's list, the probability that a listed topic is a champion is multiplied by about fifty (1.4\% to 70.8\%). The detector reaches 85 of an attainable 114.

\textbf{On 2017--2020 the detector commits zero failures in 80 slots, where chance among the candidates expected 7.6.} All three failures fall in 2021 and 2022, the vintages most affected by the COVID-19 literature (\S{}3.2). One of the three failure slots is COVID-19 itself, judged a failure because the wave passed.

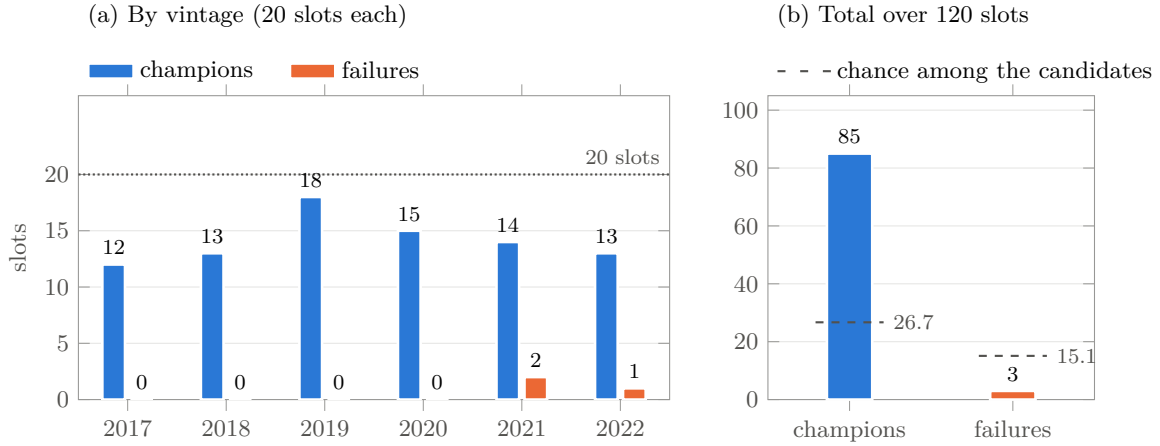
\begin{figure}[t]\centering\small
\begin{tikzpicture}
\begin{axis}[name=a, ybar, width=9.4cm, height=5.6cm, bar width=0.3cm,
  symbolic x coords={2017,2018,2019,2020,2021,2022}, xtick=data, enlarge x limits=0.1,
  ymin=0, ymax=27, ytick={0,5,10,15,20}, ylabel={slots}, ymajorgrids, grid style={ink2!15},
  axis line style={ink2!60}, tick style={ink2!60}, ticklabel style={font=\footnotesize,text=ink2},
  label style={font=\footnotesize,text=ink2}, title style={font=\footnotesize,anchor=south west,at={(0,1.14)}},
  title={(a) By vintage (20 slots each)},
  legend style={at={(0,1.01)},anchor=south west,legend columns=3,draw=none,font=\footnotesize,fill=none,
    /tikz/every even column/.append style={column sep=0.4cm}},
  nodes near coords, every node near coord/.append style={font=\scriptsize,text=black}]
\addplot[fill=cor,draw=white,line width=1pt,area legend] coordinates {(2017,12) (2018,13) (2019,18) (2020,15) (2021,14) (2022,13)};
\addplot[fill=err,draw=white,line width=1pt,area legend] coordinates {(2017,0) (2018,0) (2019,0) (2020,0) (2021,2) (2022,1)};
\legend{champions,failures}
\draw[ink2,densely dotted,thick] ({rel axis cs:0,0} |- {axis cs:2017,20}) -- ({rel axis cs:1,0} |- {axis cs:2017,20})
  node[pos=1,above left,font=\scriptsize,text=ink2]{20 slots};
\end{axis}
\begin{axis}[at={(a.south east)}, anchor=south west, xshift=1.3cm, ybar, bar shift=0pt,
  width=5.9cm, height=5.6cm, bar width=0.6cm,
  symbolic x coords={champions,failures}, xtick={champions,failures}, enlarge x limits=0.5,
  ymin=0, ymax=105, ytick={0,20,40,60,80,100}, ymajorgrids, grid style={ink2!15},
  axis line style={ink2!60}, tick style={ink2!60}, ticklabel style={font=\footnotesize,text=ink2},
  label style={font=\footnotesize,text=ink2}, title style={font=\footnotesize,anchor=south west,at={(0,1.14)}},
  title={(b) Total over 120 slots}, clip=false,
  legend style={at={(0,1.01)},anchor=south west,draw=none,font=\footnotesize,fill=none},
  nodes near coords, every node near coord/.append style={font=\scriptsize,text=black}]
\addplot[forget plot,fill=cor,draw=white,line width=1pt] coordinates {(champions,85)};
\addplot[forget plot,fill=err,draw=white,line width=1pt] coordinates {(failures,3)};
\draw[ink2,dashed,thick] ([xshift=-0.45cm]axis cs:champions,26.7) -- ([xshift=0.45cm]axis cs:champions,26.7)
  node[right,font=\scriptsize,text=ink2]{26.7};
\draw[ink2,dashed,thick] ([xshift=-0.45cm]axis cs:failures,15.1) -- ([xshift=0.45cm]axis cs:failures,15.1)
  node[right,font=\scriptsize,text=ink2]{15.1};
\addlegendimage{line legend,ink2,dashed,thick}\addlegendentry{chance among the candidates}
\end{axis}
\end{tikzpicture}
\caption{The list against chance, 2017--2022 (\S{}8.1). (a) Champion and failure slots of the detector's list, by vintage, twenty slots each: 12/0 \textperiodcentered{} 13/0 \textperiodcentered{} 18/0 \textperiodcentered{} 15/0 \textperiodcentered{} 14/2 \textperiodcentered{} 13/1. (b) Totals over the 120 slots: 85 champions and 3 failures. The dashed lines mark chance among the detector's candidates, 26.7 champions and 15.1 failures, computed as in \S{}3.3; chance is shown on the totals only.}
\label{fig:1}
\end{figure}

\subsect{8.2\texorpdfstring{\quad}{ } What has been checked}
Each line below is a control.

\begin{itemize}
\item \textbf{The verdict rule.} Nine pre-registered variants of the judge, each changing one parameter (growth multiplier, floor, horizon, denominator, failure threshold), are reported in \S{}10. The advantage holds in all nine.
\item \textbf{The date of the photograph.} Four dates were tried (31 December of \emph{Y$-$1}, 31 March, 30 June and 30 September of \emph{Y}). 30 June is the minimum of failures, not an optimum of champions.
\item \textbf{The grain of time.} With all settings unchanged and only the temporal resolution of the input degraded, the detector keeps most of its strength down to a resolution of a few weeks, when the median candidate has few visible events per period.
\item \textbf{No temporal leak in the series.} Each part of the detector that reads the dated series was recomputed with only the data prior to the photograph; the result is identical. Some other inputs were prepared after the photograph dates (Part B); that reservation stands.
\item \textbf{Two versions of one component of the specification.} The sealed file uses the earlier version, the same as all six bets. The later revision gives \textbf{the same 85/3}, with two of the 120 slots changed.
\item \textbf{Two versions of the benchmark} (sealed and rebuilt) coexist and differ by one failure. This is declared everywhere.
\item \textbf{Ten further controls} are not listed because their very statements would describe the detector. They are recorded one by one and sealed; three of them have an unfavourable or inconclusive verdict.
\end{itemize}

\subsect{8.3\texorpdfstring{\quad}{ } Outside biomedicine: illustrations, not a score}
The protocol was transposed onto OpenAlex [38], an open index of all sciences, without human designation of targets. It dates perovskite photovoltaics to 2013, neural machine translation to 2014 (the year of the founding article), binary black-hole mergers to a peak in 2016, and large language models to 2019. A mature control, Monte Carlo simulation, stays silent. These four cases are known cases chosen to be verifiable. They were not drawn from a larger set. They are illustrations, not a score.

\sect{9\texorpdfstring{\quad}{ } Places and errors: twenty against sixty}
Twenty places is a choice, not a derivation. The protocol allows the question to be asked properly: what does a longer list buy, and at what price? Before computing, we wrote and fingerprinted a short pre-registration (SHA-256, 7 October 2026; Annex B). It fixes the program (exactly that of \S{}8), the photographs (30 June, 2017--2022), the judge (\S{}3.2), and a single change: the number of places goes from twenty to sixty. Chance is computed among the same candidates and by the same method as the 26.7/15.1 of \S{}3.3. As a control, the same program re-run at twenty places in a new directory returned the sealed file of \S{}8 byte for byte.

\textbf{What it shows.} The lists are nested exactly: in every vintage, the first twenty places of the longer list are the list of \S{}8. Going from twenty to sixty places nearly doubles the share of all the candidates' champions that the list catches and lowers the share of champions among the places listed. The added places still beat chance among the same candidates on both axes, but by much less than the first twenty: the detector's advantage is concentrated at the top of its list.

\textbf{What we do not print.} The counts at sixty places, and their reference, are withheld with the specification (\S{}7.2). They are in the fingerprinted results file (Annex B). We print no count without its reference, so we print neither (Annex A, A26).

The blind spot of a twenty-place list is therefore the price of its precision. A short list favours precision; a longer list favours recall. Neither replaces the other. All bets of \S{}11 use twenty places.

\sect{10\texorpdfstring{\quad}{ } Sensitivity of the judge}
\subsect{10.1\texorpdfstring{\quad}{ } The question}
Every rate in Part B depends on our definition of a champion. An outside reader raised the objection on 9 October 2026: would another judge give another answer? Before any computation, we wrote and fingerprinted a pre-registration (SHA-256, Annex B). It fixes what does not change: the instrument and its 120 slots (the sealed file of \S{}8), the candidate population of each vintage, and the method of chance, recomputed for each variant among the same candidates exactly as for the 26.7/15.1 of \S{}3.3. In each variant \textbf{a single parameter} of the judge of \S{}3.2 changes. The counts are the same yearly counts as the benchmark, on the same material and with the same rules.

\textbf{Control, written before.} The reference judge, recomputed from these counts, had to give back exactly 85/3 and 26.7/15.1, or the measurement would stop. It did: the recomputed counts reproduce those of the benchmark for every candidate of every vintage, the labels are identical, and the list and chance return 85/3 and 26.7/15.1.

\textbf{Reading rule, written before.} The advantage ``does not depend on the definition of a champion'' if, in every variant J1 to J9, the list reaches \textbf{at least twice} the champions of chance in the same variant \textbf{and has no more failures} than chance. Otherwise each failing variant is named in this paper as a limitation.

\needlines{18}
\subsect{10.2\texorpdfstring{\quad}{ } The result}
The reference first, then the variants:

\nopagebreak[4]
\begin{center}\small
\begin{tabularx}{\textwidth}{@{}P{2.7cm}YR{2.0cm}R{2.0cm}R{1.8cm}l@{}}
\toprule
variant & what changes in the judge & list, champions / failures & chance, champions / failures & champions, ratio to chance & rule \\
\midrule
J0 (reference, \S{}3.2) & --- & \textbf{85 / 3} & 26.7 / 15.1 & 3.2 & control \\
J1 & growth $\times$2 instead of $\times$3 & 99 / 3 & 32.5 / 15.1 & 3.0 & met \\
J2 & growth $\times$4 & 65 / 3 & 17.9 / 15.1 & 3.6 & met \\
J3 & floor of 50 records instead of 100 & 86 / 3 & 30.9 / 15.1 & 2.8 & met \\
J4 & floor of 200 records & 57 / 3 & 11.3 / 15.1 & 5.0 & met \\
J5 & horizon \emph{Y}+2 instead of \emph{Y}+3 & 65 / 2 & 16.2 / 13.1 & 4.0 & met \\
J6 & horizon \emph{Y}+4 (vintages 2017--2021, 100 slots) & 67 / 2 & 26.9 / 12.5 & 2.5 & met \\
J7 & denominator \emph{n(Y$-$2)} instead of \emph{n(Y$-$1)} & 102 / 0 & 33.3 / 6.1 & 3.1 & met \\
J8 & failure at 0.5 times instead of 0.7 & 85 / 2 & 26.7 / 9.5 & 3.2 & met \\
J9 & failure at 0.8 times & 85 / 3 & 26.7 / 19.8 & 3.2 & met \\
\bottomrule
\end{tabularx}
\end{center}

\emph{120 slots in every line except J6. Chance counts champions as distinct topics after merging equivalent names, as in \S{}3.3. J6 stops at the 2021 vintage, since its verdict for 2022 would need a complete 2026; on the same five vintages, the reference judge gives a ratio of 3.1.}

\textbf{By the rule written before, the advantage does not depend on the definition of a champion.} All nine variants meet both conditions. No variant fails, so none is named as a limitation.

\begin{figure}[t]\centering\small
\begin{tikzpicture}
\begin{axis}[name=a, ybar, bar shift=0pt, width=8.3cm, height=5.6cm, bar width=0.4cm,
  symbolic x coords={J0,J1,J2,J3,J4,J5,J6,J7,J8,J9}, xtick={J0,J1,J2,J3,J4,J5,J6,J7,J8,J9}, enlarge x limits=0.07,
  ymin=0, ymax=5.8, ylabel={ratio to chance}, ymajorgrids, grid style={ink2!15},
  axis line style={ink2!60}, tick style={ink2!60}, ticklabel style={font=\footnotesize,text=ink2},
  label style={font=\footnotesize,text=ink2}, title style={font=\footnotesize,anchor=south west,at={(0,1.14)}},
  title={(a) Champions: ratio of the list to chance},
  legend style={at={(0,1.01)},anchor=south west,draw=none,font=\footnotesize,fill=none},
  nodes near coords, every node near coord/.append style={font=\scriptsize,text=black,/pgf/number format/.cd,fixed,fixed zerofill,precision=1}]
\addplot[forget plot,fill=cor,draw=white,line width=1pt] coordinates {(J0,3.2)};
\addplot[forget plot,fill=cor!45,draw=white,line width=1pt] coordinates {(J1,3.0) (J2,3.6) (J3,2.8) (J4,5.0) (J5,4.0) (J6,2.5) (J7,3.1) (J8,3.2) (J9,3.2)};
\draw[ink2,dashed,thick] ({rel axis cs:0,0} |- {axis cs:J0,2}) -- ({rel axis cs:1,0} |- {axis cs:J0,2});
\addlegendimage{line legend,ink2,dashed,thick}\addlegendentry{threshold written before ($\times$2)}
\end{axis}
\begin{axis}[at={(a.outer east)}, anchor=outer west, xshift=0.1cm, ybar, width=8.3cm, height=5.6cm, bar width=0.17cm,
  symbolic x coords={J0,J1,J2,J3,J4,J5,J6,J7,J8,J9}, xtick=data, enlarge x limits=0.07,
  ymin=0, ymax=24, ytick={0,5,10,15,20}, ylabel={failure slots}, ymajorgrids, grid style={ink2!15},
  axis line style={ink2!60}, tick style={ink2!60}, ticklabel style={font=\footnotesize,text=ink2},
  label style={font=\footnotesize,text=ink2}, title style={font=\footnotesize,anchor=south west,at={(0,1.14)}},
  title={(b) Failures: the list against chance},
  legend style={at={(0,1.01)},anchor=south west,legend columns=2,draw=none,font=\footnotesize,fill=none,
    /tikz/every even column/.append style={column sep=0.4cm}}]
\addplot[area legend,fill=err,draw=err,nodes near coords,every node near coord/.append style={font=\scriptsize,text=black}] coordinates {(J0,3) (J1,3) (J2,3) (J3,3) (J4,3) (J5,2) (J6,2) (J7,0) (J8,2) (J9,3)};
\addplot[area legend,pattern=north east lines,pattern color=err,draw=err] coordinates {(J0,15.1) (J1,15.1) (J2,15.1) (J3,15.1) (J4,15.1) (J5,13.1) (J6,12.5) (J7,6.1) (J8,9.5) (J9,19.8)};
\legend{the list,chance among the candidates}
\end{axis}
\end{tikzpicture}
\caption{Sensitivity of the judge (\S{}10.2). (a) Ratio of the list's champions to those of chance among the same candidates, for the reference judge J0 (dark bar) and the nine pre-registered variants J1 to J9; the dashed line is the threshold of two written before computing. (b) Failure slots of the list (solid) and of chance (hatched) in each variant. 120 slots in every variant except J6 (vintages 2017--2021, 100 slots). All nine variants meet both conditions of the rule.}
\label{fig:2}
\end{figure}
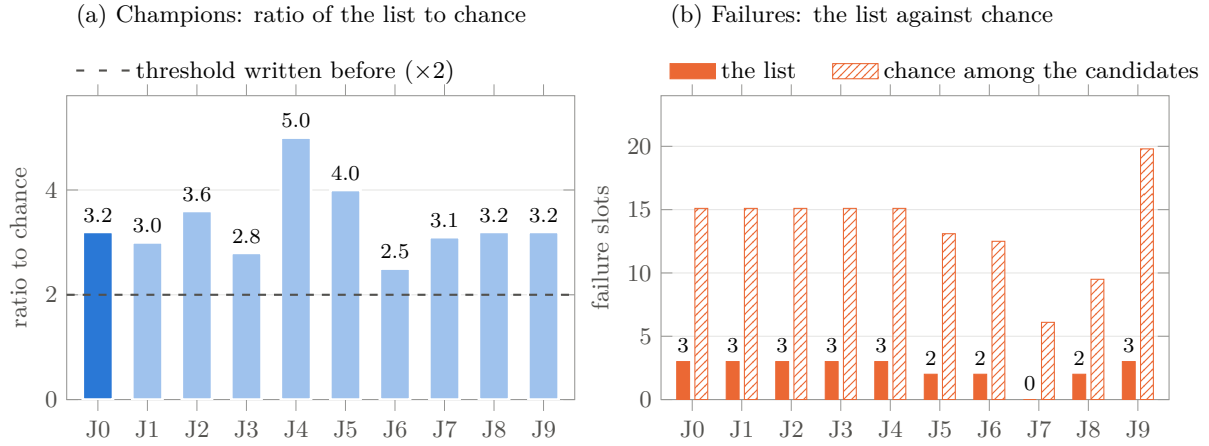

\subsect{10.3\texorpdfstring{\quad}{ } Reading after the fact}
These three remarks were not pre-registered.

\begin{itemize}
\item \textbf{The longest horizon is the weakest variant} (J6: 2.5 times chance, against 3.1 for the reference judge on the same five vintages). Four years after the photograph, more of the topics drawn at random cross the bar, and the gap narrows. It stays above the threshold of two.
\item \textbf{The list's failures barely move} (0 to 3 in every variant), while those of chance range from 6 to 20. On the failure axis, the size of the gain depends on the threshold more than on the list.
\item \textbf{J7 moves the denominator one year away} from the year that immediately precedes the photograph (\S{}3.10). The list then has no failure at all.
\end{itemize}

These results bear on the retrospective benchmark. They say nothing about the sealed bets.

\sect{Part C --- Prospective validation: the sealed bets}
\begin{quote}
\textbf{Which bets test which instrument.} Bets 1 to 4 \textbf{do not test exactly} the instrument that produced the 85/3 of \S{}8: each differs from it in parts of the specification, and bet 2 carries a revision whose validation read the future (\S{}6.4). \textbf{Bet 5 is the first bet that tests that exact instrument.} \textbf{Bet 6 tests it on up-to-date data}: the photograph of 30 June 2026, read on the baseline plus the updates to 6 October 2026.
\end{quote}

Sealing proves that a list existed at a date, not that it is right. The verdicts will say whether they are.

\sect{11\texorpdfstring{\quad}{ } Sealed bets and dates of judgement}
\subsect{11.1\texorpdfstring{\quad}{ } Why bets}
Everything above is retrospective (Part B). A sealed bet is not: its outcome is unknown to us as to the reader. Of the six bets, bets 5 and 6 alone test exactly the instrument of the benchmark, and claims about future performance rest on them.

\subsect{11.2\texorpdfstring{\quad}{ } The six bets}
\nopagebreak[4]
\begin{center}\small
\begin{tabularx}{\textwidth}{@{}lllYP{4.6cm}@{}}
\toprule
bet & photograph & sealed & what it tests & verdict \\
\midrule
1 & 30-06-2025 & 01-09-2026 & the frozen protocol of 31 August 2026 & verdict date 31-12-2028 (sealed); read on complete 2028 volumes, early 2029 (A25) \\
2 & 30-06-2025 & 01-09-2026 & a revised instrument & same \\
3 & 30-06-2025 & 01-09-2026 & a third revision, one change from bet 2 & same \\
4 & 30-06-2025 & 01-09-2026 & the full score, one change from bet 3 & same \\
5 & 30-06-2025 & 07-10-2026 & \textbf{exactly the program of the 85/3 benchmark} & same \\
6 & 30-06-2026 & 07-10-2026 & the same program, on the baseline plus updates to 6 October 2026 & judged on 2029 volumes, read once the indexing of 2029 is complete, early 2030 (A25) \\
\bottomrule
\end{tabularx}
\end{center}

All bets list \textbf{twenty places} and are judged by the rule of \S{}3.2: for vintage 2025, \emph{n(2028)} against \emph{n(2024)}; for vintage 2026, \emph{n(2029)} against \emph{n(2025)}. Their fingerprints are in Annex B.

\textbf{Bets 1 to 4.} Four lists of vintage 2025, sealed on 1 September 2026, fourteen months after their photograph. Each tests one version of the instrument: the frozen protocol, then three successive revisions, each isolating a single change, on the same vintage and before the same judge. Earlier versions of these lists, sealed from 25 August to 1 September, held the same topic in more than one position. They were \textbf{withdrawn and replaced before any circulation}; each withdrawal is declared in an anchored manifest (Annex A, A8). The current versions are those of manifest v19.

An audit made on 7 October 2026 established a point that we state here plainly: \textbf{none of the four lists is exactly the instrument of the benchmark table of \S{}8.} They differ from it in parts of the specification. Bet 2, moreover, carries the revision whose validation read the future (\S{}6.4).

\textbf{Bet 5.} Sealed on 7 October 2026, on the same photograph of 30 June 2025. It applies \textbf{exactly the program that produced the 85/3}, re-run byte for byte the same morning (\S{}7.1), with no change other than the list of vintages and the output paths. It is the first bet that tests the instrument of the published benchmark. Its pre-registration was written and timestamped before any computation of the photograph. Two amendments were written and fingerprinted before sealing, both of a technical nature (a memory limit and an internal control); neither changes how the list is ranked. The delay between photograph and sealing is \textbf{fifteen months and one week}. During that delay, the year 2025 was entirely written and read by us. The program reads nothing after June 2025, but the substrate is the 2026 baseline, one input of the program was prepared after the photograph, and we knew the lists of bets 1 to 4. No choice was made: the program is frozen, and its six files are fingerprinted (Annex B).

\textbf{Bet 6.} Sealed on 7 October 2026, on the photograph of 30 June 2026. The 2026 baseline stops in practice in January 2026, so the bet reads the baseline \textbf{plus the update files to 6 October 2026} (\S{}4.1). June 2026 was judged complete enough: 152,290 records, 1.18 times June 2025 at the same delay. The pre-registration \textbf{declares an amendment before computing}: with the updates, the 2025 photograph gives the same twenty topics as bet 5 in a different order (eight positions change), the candidate population gains two topics, and an internal control changes by one champion. The cause is late records and revisions. Bet 5 stays unchanged and valid on the baseline alone. The delay between photograph and sealing is \textbf{three months and one week}.

\subsect{11.3\texorpdfstring{\quad}{ } Chance and rate for twenty places}
Chance is computed \textbf{at the verdict}, among the sealed candidates of each bet: 20 $\times$ champions / candidates, and the same for failures. The historical reference, from the benchmark, is \textbf{4.45 champions and 2.52 failures for twenty places} (rounded to 4.5 and 2.5 in our earlier public evidence file). The rate of the benchmark is \textbf{14.2 champions and 0.5 failure for twenty places}. The three outcomes of \S{}3.9 are read against these two references.

\subsect{11.4\texorpdfstring{\quad}{ } How to check}
The anchored objects are the manifests v19, v53 and v54, whose fingerprints are in Annex B. Their OpenTimestamps proofs, and the public files of the twenty position fingerprints of bets 5 and 6, are deposited at doi:\url{10.5281/zenodo.23641352} (Zenodo). Anyone can check with \texttt{ots verify} that each manifest existed at the date stated. Checking that a bet's fingerprint appears in its manifest requires the manifest, which also names internal files; it is shown under agreement. The proof of bets 1 to 4 (manifest v19) is anchored in Bitcoin blocks 965063, 965072 and 965109, the earliest mined on 1 September 2026 at 17:33 UTC. The proof of manifest v53, which declares bet 5, is anchored in block 970307, mined on 7 October 2026 at 07:24 UTC. The proof of manifest v54, which declares bet 6, is anchored in blocks 970310, 970332 and 970333. All six bets are now anchored. At the verdict, the lists and the salts of each position will be revealed, so that every position fingerprint can be recomputed; the chance values computed on the fingerprinted candidate populations will be published.

\needlines{13}
\subsect{11.5\texorpdfstring{\quad}{ } Datings already judged against the nomenclature}
A dating answers a different question from a bet: given a name, since when has the topic been rising? It is back-computed under frozen rules on the corpus prior to each date. A register of six positions was fingerprinted on 25 August 2026 (file \texttt{8f3b7815\dots{}}), cited in a document anchored in the Bitcoin chain on 26 August 2026. Five positions have been opened. We judge them against the slowest judge there is, the creation of the official MeSH descriptor:

\nopagebreak[4]
\begin{center}\small
\begin{tabular}{@{}lllr@{}}
\toprule
position revealed & dating (first signal, back-computed) & descriptor created & lead \\
\midrule
spatial transcriptomics & December 2018 & January 2026 & 7.1 years \\
federated learning & September 2020 & January 2025 & 4.3 years \\
graph neural networks & December 2020 & January 2025 & 4.1 years \\
tirzepatide & December 2020 & January 2025 & 4.1 years \\
long covid & January 2022 & January 2023 & 12 months \\
\bottomrule
\end{tabular}
\end{center}

A sixth position, whose descriptor predates the dating, is a \textbf{lost bet}. \textbf{These datings are back-computed.} The public sealing dates from 2026. That is exactly the limit the bets of \S{}11.2 lift. Two further names remain in plain view, with judgement to come: lactylation (dating end of 2021; 608 records in 2025; horizon 2029) and cardiovascular-kidney-metabolic syndrome (dating summer 2024; 362 records in 2025; horizon 2030).

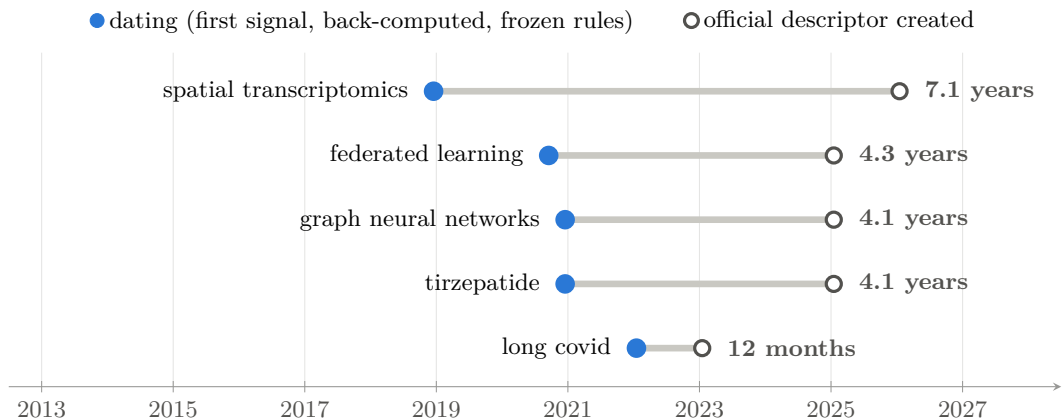
\begin{figure}[t]\centering\small
\begin{tikzpicture}
\begin{axis}[width=15.5cm, height=6cm, xmin=2012.5, xmax=2028.5, ymin=0.4, ymax=5.6, y dir=reverse,
  axis x line=bottom, axis y line=none, xtick={2013,2015,...,2027}, xticklabel style={/pgf/number format/1000 sep={}},
  xmajorgrids, grid style={ink2!15}, clip=false,
  axis line style={ink2!60}, tick style={ink2!60}, ticklabel style={font=\footnotesize,text=ink2},
  label style={font=\footnotesize,text=ink2}, title style={font=\footnotesize,anchor=south west,at={(0,1.14)}},
  legend style={at={(0.5,1.03)},anchor=south,legend columns=2,draw=none,font=\footnotesize,/tikz/every even column/.append style={column sep=0.6cm}}]
\addlegendimage{only marks,mark=*,mark size=2.6pt,mark options={fill=cor,draw=cor}}
\addlegendentry{dating (first signal, back-computed, frozen rules)}
\addlegendimage{only marks,mark=o,mark size=2.6pt,mark options={draw=ink2,line width=1.2pt}}
\addlegendentry{official descriptor created}
\draw[nei,line width=2.5pt] (axis cs:2018.958,1) -- (axis cs:2026.042,1);
\node[circle,fill=cor,inner sep=2.6pt] at (axis cs:2018.958,1) {};
\node[circle,draw=ink2,fill=white,line width=1.2pt,inner sep=2pt] at (axis cs:2026.042,1) {};
\node[anchor=east,font=\footnotesize] at ([xshift=-6pt]axis cs:2018.958,1) {spatial transcriptomics};
\node[anchor=west,font=\footnotesize\bfseries,text=ink2] at ([xshift=6pt]axis cs:2026.042,1) {7.1 years};
\draw[nei,line width=2.5pt] (axis cs:2020.708,2) -- (axis cs:2025.042,2);
\node[circle,fill=cor,inner sep=2.6pt] at (axis cs:2020.708,2) {};
\node[circle,draw=ink2,fill=white,line width=1.2pt,inner sep=2pt] at (axis cs:2025.042,2) {};
\node[anchor=east,font=\footnotesize] at ([xshift=-6pt]axis cs:2020.708,2) {federated learning};
\node[anchor=west,font=\footnotesize\bfseries,text=ink2] at ([xshift=6pt]axis cs:2025.042,2) {4.3 years};
\draw[nei,line width=2.5pt] (axis cs:2020.958,3) -- (axis cs:2025.042,3);
\node[circle,fill=cor,inner sep=2.6pt] at (axis cs:2020.958,3) {};
\node[circle,draw=ink2,fill=white,line width=1.2pt,inner sep=2pt] at (axis cs:2025.042,3) {};
\node[anchor=east,font=\footnotesize] at ([xshift=-6pt]axis cs:2020.958,3) {graph neural networks};
\node[anchor=west,font=\footnotesize\bfseries,text=ink2] at ([xshift=6pt]axis cs:2025.042,3) {4.1 years};
\draw[nei,line width=2.5pt] (axis cs:2020.958,4) -- (axis cs:2025.042,4);
\node[circle,fill=cor,inner sep=2.6pt] at (axis cs:2020.958,4) {};
\node[circle,draw=ink2,fill=white,line width=1.2pt,inner sep=2pt] at (axis cs:2025.042,4) {};
\node[anchor=east,font=\footnotesize] at ([xshift=-6pt]axis cs:2020.958,4) {tirzepatide};
\node[anchor=west,font=\footnotesize\bfseries,text=ink2] at ([xshift=6pt]axis cs:2025.042,4) {4.1 years};
\draw[nei,line width=2.5pt] (axis cs:2022.042,5) -- (axis cs:2023.042,5);
\node[circle,fill=cor,inner sep=2.6pt] at (axis cs:2022.042,5) {};
\node[circle,draw=ink2,fill=white,line width=1.2pt,inner sep=2pt] at (axis cs:2023.042,5) {};
\node[anchor=east,font=\footnotesize] at ([xshift=-6pt]axis cs:2022.042,5) {long covid};
\node[anchor=west,font=\footnotesize\bfseries,text=ink2] at ([xshift=6pt]axis cs:2023.042,5) {12 months};
\end{axis}
\end{tikzpicture}
\caption{The five opened positions of the register fingerprinted on 25 August 2026, judged against the creation of the official MeSH descriptor (\S{}11.5). Filled circle: dating (first signal, back-computed under frozen rules); open circle: descriptor created. Leads of 7.1, 4.3, 4.1 and 4.1 years, and 12 months. These datings are back-computed; the public sealing dates from 2026.}
\label{fig:3}
\end{figure}

\sect{12\texorpdfstring{\quad}{ } A secondary, exploratory study: research funders}
\subsect{12.1\texorpdfstring{\quad}{ } The question}
Funders finance emerging fields early, often before the evidence is settled. Two questions follow. \textbf{How early does the money arrive} on the topics that later take off? \textbf{How well does it choose?} The first is often asked. The second is rarely measured, because it needs a fixed date, a fixed list of alternatives and a verdict written in advance. The protocol of \S{}3 provides all three.

\textbf{This study is secondary and exploratory.} It is neither a validation of the detector nor an evaluation of the funders. Its pre-registrations were fingerprinted but not externally timestamped. Its matching of topics to grant texts is automatic and \textbf{has not been audited by hand}: a blind manual audit of a sample of the matches, in which a reader sees the grant and the topic but not the verdict, and says whether the grant is about the topic, is future work. Rates are read with their 95\% intervals where the groups are small enough for these to matter (\S{}12.4).

We measured five funders first (3 October 2026). A third addendum, written and fingerprinted before any download, then added three funders in the United States and a corrected dating of NIH grants (8 October 2026). Everything below is reported under the same rules for all of them.

\needlines{17}
\subsect{12.2\texorpdfstring{\quad}{ } Data and method}
Everything on the funding side is public and reproducible without us.

\nopagebreak[4]
\begin{center}\small
\begin{tabularx}{\textwidth}{@{}P{3.3cm}YR{2.7cm}P{2.6cm}R{1.5cm}@{}}
\toprule
funder & source & grants 2010--2025 & date used & with long text \\
\midrule
Wellcome & 360Giving grants file (to 4 August 2026) & 15,151 & award date & 86\% \\
NIH, record by record & NIH ExPORTER, annual records and abstracts & 1,202,797 records & the record's own budget start & 96\% \\
NIH, by project (first variant) & NIH ExPORTER, projects and abstracts & 244,160 projects & first project start & 96\% \\
UKRI & OpenAIRE Graph, projects & 149,018 & start date & 92\% \\
European Commission & OpenAIRE Graph, projects & 72,893 & start date & 73\% \\
ANR (France) & OpenAIRE Graph, projects & 26,858 & start date & 87\% \\
NSF & NSF Award Search, yearly files & 193,757 & award start date & 99\% \\
Gates Foundation (titles only) & grants file published by the foundation (to 1 October 2026) & 28,223 & month committed & 21\% \\
Chan Zuckerberg Initiative (titles only) & grants list published by the foundation & 5,503 (2018--2024 only) & commitment year & 26\% \\
\bottomrule
\end{tabularx}
\end{center}

\emph{``Long text'': a title and description of at least 200 characters.}

\begin{itemize}
\item \textbf{Matching.} A grant names a topic when the topic's name appears as a whole expression in its title or description, lower-cased, with punctuation removed; the last word may take a plural ``s''. No synonyms and no language model. The same code, re-run on the first five funders, reproduces their published results exactly.
\item \textbf{``Funded'' at a date.} Here a topic is ``funded'' by a funder when at least one of its grants names the topic, dated from 1 January 2010 to the day before the photograph. ``Most funded'' means named in the most grants. Award amounts are not used.
\item \textbf{Units.} Rates are counted in candidate-years, a topic at a given June. Timing figures count distinct topics.
\item \textbf{NIH, record by record.} Each annual NIH record is dated by its own budget start and read with its own title and abstract only; a record counts as one grant. This corrects our first NIH variant (\S{}12.6). Records with no budget start (5.8\%) are left out. Both variants are reported.
\item \textbf{Titles only.} The Gates Foundation and the Chan Zuckerberg Initiative (CZI) publish one sentence of purpose per grant, with no abstract. They are measured and marked ``titles only'': theirs is the coarsest comparison. CZI's published grants start in 2018, so it could fund nothing before the 2017 and 2018 photographs.
\item \textbf{Coarse dates.} Gates publishes a month, which we set on the 15th; the first or last day of the month gives identical results. CZI publishes a year, which we set on 1 July; 31 December gives the same results, and 1 January moves its champion rate by about one point.
\item \textbf{Admission.} The third addendum admitted a funder only if it had at least 1,000 grants dated 2010--2025 with a text, at least half of them with 50 characters or more. The Burroughs Wellcome Fund (which publishes awardees and institutions but no project titles) and the Howard Hughes Medical Institute (whose website refused automated reading of its public pages, which we did not bypass) could not be measured, and no figure is given for them.
\item \textbf{Former names.} Before the results of the US funders were read, we wrote a table of the other forms a grant text may use for each of the 63 champion names of the list: former names, long forms of acronyms, British forms and drug code names. It was applied to every funder, and it governs two statements of \S{}12.6: whether the list came before a funder, and whether a champion was never funded.
\item \textbf{Protocol.} The main protocol and three addenda were written and fingerprinted (SHA-256) before the computation each governs. \textbf{These fingerprints were not externally timestamped.} They are listed in Annex B.
\item \textbf{Reading rules, written before.} Money ``chooses better than the population'' if its champion rate among funded candidates reaches at least 1.5 times the population's. The list ``chooses better than the money'' if its champion rate exceeds the funder's by at least 20 points, with no more failures in proportion. A lead of the list over a funder may be claimed only if at least ten listed champions were first named by that funder after the list.
\end{itemize}

The reference throughout is the candidate population of the benchmark: \textbf{20.2\% of candidate-years became champions and 13.0\% failures.} Weighted like the list (twenty slots a year), chance gives 22.3\% and 12.6\%. The list is the detector's 120 slots of \S{}8: 70.8\% champions and 2.5\% failures.

\subsect{12.3\texorpdfstring{\quad}{ } Funding chooses broadly}
For the large public portfolios, the 95\% Wilson interval on the champion rate is narrow. NIH's interval includes the population rate under both datings; Wellcome's, UKRI's, the Commission's, ANR's and NSF's lie just above it. Gates's interval includes the population rate. CZI's is wide (22.7--49.2\%), on a small sample.

\needlines{15}
Among all candidate topics already funded by each funder at the photograph, that is, named in at least one of its grants:

\nopagebreak[4]
\begin{center}\small
\begin{tabular}{@{}lrrr@{}}
\toprule
already funded by & champions & failures & ratio to the population \\
\midrule
Wellcome & 25.5\% & 10.0\% & 1.26 \\
NIH, record by record & 20.8\% & 12.3\% & 1.03 \\
NIH, by project (first variant) & 21.5\% & 11.9\% & 1.07 \\
UKRI & 24.7\% & 10.4\% & 1.22 \\
European Commission & 27.0\% & 9.9\% & 1.34 \\
ANR & 25.2\% & 10.6\% & 1.25 \\
NSF & 24.9\% & 11.1\% & 1.24 \\
Gates Foundation (titles only) & 22.8\% & 18.6\% & 1.13 \\
CZI (titles only) & 34.8\% & 21.7\% & 1.73 \\
\emph{population} & \emph{20.2\%} & \emph{13.0\%} & \emph{1.00} \\
\bottomrule
\end{tabular}
\end{center}

\textbf{Only one funder reaches 1.5 times the population} in the main variant: CZI (1.73). The rule says yes, and we report it as such, with four reservations. Its published grants start in 2018, which makes its sample the smallest of the study by far. Most of its champions belong to one family, single-cell biology and spatial transcriptomics: a successful thematic bet, not a general selectivity. One of its champions is a false match, a topic name that matched a grant to a county of the same name; without it, CZI stands at 33\% (1.65 times). And more of its funded topics became failures than in the population (21.7\% against 13.0\%): COVID-19-era topics named in grants of its education and community programmes, outside its science programme. Two secondary variants of the first study also cross the bar (Wellcome on grants of the last 24 months, 1.51; the Commission with at least three grants, 1.53); the rule applies to the main variant and gives no sentence for them.

The large public portfolios touch a large share of all candidates and pick only slightly better than chance. Gates touches far fewer, and the topics it does touch failed more often than the population (18.6\% against 13.0\%); this describes the literature, not the grants. \textbf{The list exceeds every funder by 36 to 50 points, with fewer failures} (2.5\% against 9.9\% to 21.7\%).

\needlines{19}
\subsect{12.4\texorpdfstring{\quad}{ } At equal size}
A list of twenty is narrower than a funder's whole portfolio. An internal review (by an AI agent) pointed out that the comparison above is therefore unfair to funders. We wrote a first addendum, fingerprinted it before computing, and took each funder's up to twenty candidate topics named in the most grants each year. Intervals are 95\% Wilson intervals, and they come first: with groups of this size, differences below about ten points are within noise.

\nopagebreak[4]
\begin{center}\small
\begin{tabularx}{\textwidth}{@{}YR{0.9cm}R{3.2cm}R{3.0cm}@{}}
\toprule
up to twenty topics a year, 2017--2022, chosen by & slots & champions [95\% interval] & failures [95\% interval] \\
\midrule
chance, weighted like the list & 120 & 22.3\% & 12.6\% \\
Wellcome & 111 & 28.8\% [21.2--37.9] & 15.3\% [9.8--23.2] \\
NIH, record by record & 120 & 20.8\% [14.5--28.9] & 14.2\% [9.0--21.5] \\
NIH, by project (first variant) & 120 & 25.8\% [18.8--34.3] & 14.2\% [9.0--21.5] \\
UKRI & 120 & 35.8\% [27.8--44.7] & 14.2\% [9.0--21.5] \\
European Commission & 120 & 35.8\% [27.8--44.7] & 10.0\% [5.8--16.7] \\
ANR & 113 & 27.4\% [20.1--36.3] & 20.4\% [14.0--28.7] \\
NSF & 120 & 25.8\% [18.8--34.3] & 16.7\% [11.1--24.3] \\
Gates Foundation (titles only) & 106 & 20.8\% [14.1--29.4] & 18.9\% [12.6--27.4] \\
CZI (titles only) & 46 & 34.8\% [22.7--49.2] & 21.7\% [12.3--35.6] \\
\textbf{the detector} & \textbf{120} & \textbf{70.8\% [62.2--78.2]} & \textbf{2.5\% [0.9--7.1]} \\
\bottomrule
\end{tabularx}
\end{center}

\emph{95\% Wilson intervals. CZI names fewer than twenty candidate topics a year, so its line holds all the topics it named.}

On grants of the last 24 months only, the funders' top twenty range from 20.0\% (Gates) to 44.2\% (UKRI) champions (Wellcome 34.5\%). The rule ``at equal size, the list chooses better'' holds for all eight funders, under both NIH datings. Two of them, NIH dated record by record and Gates, sit below chance weighted like the list (20.8\% against 22.3\%), well within noise. Wellcome's top twenty share between one and five topics a year with the detector's list.

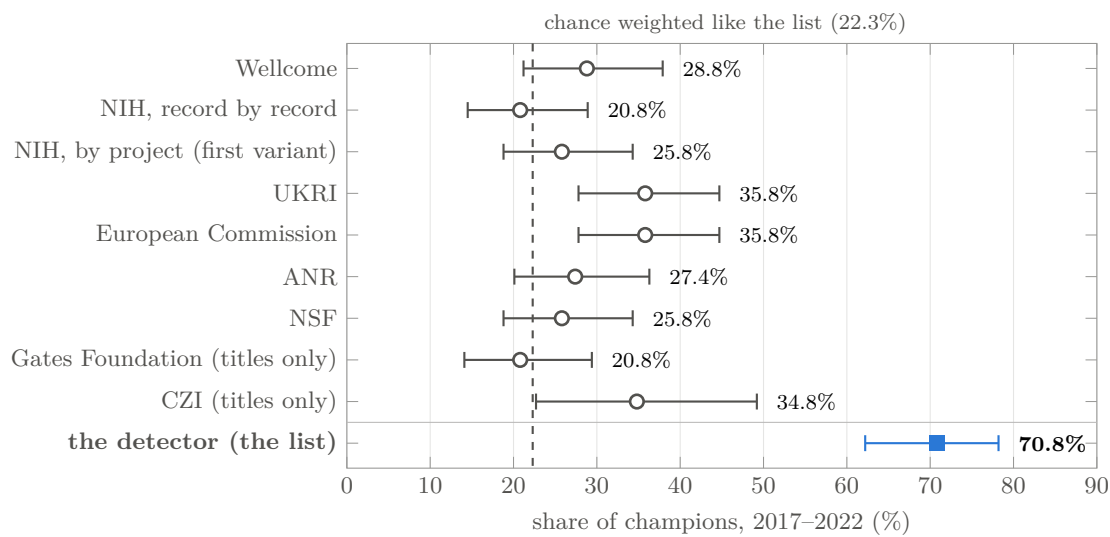
\begin{figure}[t]\centering\small
\begin{tikzpicture}
\begin{axis}[width=11.5cm, height=7.2cm, xmin=0, xmax=90, ymin=0.4, ymax=10.6, y dir=reverse,
  ytick={1,...,10}, yticklabels={{Wellcome},{NIH, record by record},{NIH, by project (first variant)},{UKRI},{European Commission},{ANR},{NSF},{Gates Foundation (titles only)},{CZI (titles only)},{\textbf{the detector (the list)}}},
  xlabel={share of champions, 2017--2022 (\%)}, xmajorgrids, grid style={ink2!15}, clip=false,
  axis line style={ink2!60}, tick style={ink2!60}, ticklabel style={font=\footnotesize,text=ink2},
  label style={font=\footnotesize,text=ink2}, title style={font=\footnotesize,anchor=south west,at={(0,1.14)}},
  ]
\draw[ink2!40] ({rel axis cs:0,0} |- {axis cs:0,9.5}) -- ({rel axis cs:1,0} |- {axis cs:0,9.5});
\draw[ink2,dashed,thick] (axis cs:22.3,0.4) -- (axis cs:22.3,10.6);
\node[anchor=south west,font=\scriptsize,text=ink2] at (axis cs:22.6,0.4) {chance weighted like the list (22.3\%)};
\draw[ink2,thick] (axis cs:21.2,1) -- (axis cs:37.9,1);
\draw[ink2,thick] ([yshift=-3pt]axis cs:21.2,1) -- ([yshift=3pt]axis cs:21.2,1);
\draw[ink2,thick] ([yshift=-3pt]axis cs:37.9,1) -- ([yshift=3pt]axis cs:37.9,1);
\node[circle,draw=ink2,fill=white,line width=1pt,inner sep=1.8pt] at (axis cs:28.8,1) {};
\node[anchor=west,font=\scriptsize] at ([xshift=4pt]axis cs:37.9,1) {28.8\%};
\draw[ink2,thick] (axis cs:14.5,2) -- (axis cs:28.9,2);
\draw[ink2,thick] ([yshift=-3pt]axis cs:14.5,2) -- ([yshift=3pt]axis cs:14.5,2);
\draw[ink2,thick] ([yshift=-3pt]axis cs:28.9,2) -- ([yshift=3pt]axis cs:28.9,2);
\node[circle,draw=ink2,fill=white,line width=1pt,inner sep=1.8pt] at (axis cs:20.8,2) {};
\node[anchor=west,font=\scriptsize] at ([xshift=4pt]axis cs:28.9,2) {20.8\%};
\draw[ink2,thick] (axis cs:18.8,3) -- (axis cs:34.3,3);
\draw[ink2,thick] ([yshift=-3pt]axis cs:18.8,3) -- ([yshift=3pt]axis cs:18.8,3);
\draw[ink2,thick] ([yshift=-3pt]axis cs:34.3,3) -- ([yshift=3pt]axis cs:34.3,3);
\node[circle,draw=ink2,fill=white,line width=1pt,inner sep=1.8pt] at (axis cs:25.8,3) {};
\node[anchor=west,font=\scriptsize] at ([xshift=4pt]axis cs:34.3,3) {25.8\%};
\draw[ink2,thick] (axis cs:27.8,4) -- (axis cs:44.7,4);
\draw[ink2,thick] ([yshift=-3pt]axis cs:27.8,4) -- ([yshift=3pt]axis cs:27.8,4);
\draw[ink2,thick] ([yshift=-3pt]axis cs:44.7,4) -- ([yshift=3pt]axis cs:44.7,4);
\node[circle,draw=ink2,fill=white,line width=1pt,inner sep=1.8pt] at (axis cs:35.8,4) {};
\node[anchor=west,font=\scriptsize] at ([xshift=4pt]axis cs:44.7,4) {35.8\%};
\draw[ink2,thick] (axis cs:27.8,5) -- (axis cs:44.7,5);
\draw[ink2,thick] ([yshift=-3pt]axis cs:27.8,5) -- ([yshift=3pt]axis cs:27.8,5);
\draw[ink2,thick] ([yshift=-3pt]axis cs:44.7,5) -- ([yshift=3pt]axis cs:44.7,5);
\node[circle,draw=ink2,fill=white,line width=1pt,inner sep=1.8pt] at (axis cs:35.8,5) {};
\node[anchor=west,font=\scriptsize] at ([xshift=4pt]axis cs:44.7,5) {35.8\%};
\draw[ink2,thick] (axis cs:20.1,6) -- (axis cs:36.3,6);
\draw[ink2,thick] ([yshift=-3pt]axis cs:20.1,6) -- ([yshift=3pt]axis cs:20.1,6);
\draw[ink2,thick] ([yshift=-3pt]axis cs:36.3,6) -- ([yshift=3pt]axis cs:36.3,6);
\node[circle,draw=ink2,fill=white,line width=1pt,inner sep=1.8pt] at (axis cs:27.4,6) {};
\node[anchor=west,font=\scriptsize] at ([xshift=4pt]axis cs:36.3,6) {27.4\%};
\draw[ink2,thick] (axis cs:18.8,7) -- (axis cs:34.3,7);
\draw[ink2,thick] ([yshift=-3pt]axis cs:18.8,7) -- ([yshift=3pt]axis cs:18.8,7);
\draw[ink2,thick] ([yshift=-3pt]axis cs:34.3,7) -- ([yshift=3pt]axis cs:34.3,7);
\node[circle,draw=ink2,fill=white,line width=1pt,inner sep=1.8pt] at (axis cs:25.8,7) {};
\node[anchor=west,font=\scriptsize] at ([xshift=4pt]axis cs:34.3,7) {25.8\%};
\draw[ink2,thick] (axis cs:14.1,8) -- (axis cs:29.4,8);
\draw[ink2,thick] ([yshift=-3pt]axis cs:14.1,8) -- ([yshift=3pt]axis cs:14.1,8);
\draw[ink2,thick] ([yshift=-3pt]axis cs:29.4,8) -- ([yshift=3pt]axis cs:29.4,8);
\node[circle,draw=ink2,fill=white,line width=1pt,inner sep=1.8pt] at (axis cs:20.8,8) {};
\node[anchor=west,font=\scriptsize] at ([xshift=4pt]axis cs:29.4,8) {20.8\%};
\draw[ink2,thick] (axis cs:22.7,9) -- (axis cs:49.2,9);
\draw[ink2,thick] ([yshift=-3pt]axis cs:22.7,9) -- ([yshift=3pt]axis cs:22.7,9);
\draw[ink2,thick] ([yshift=-3pt]axis cs:49.2,9) -- ([yshift=3pt]axis cs:49.2,9);
\node[circle,draw=ink2,fill=white,line width=1pt,inner sep=1.8pt] at (axis cs:34.8,9) {};
\node[anchor=west,font=\scriptsize] at ([xshift=4pt]axis cs:49.2,9) {34.8\%};
\draw[cor,thick] (axis cs:62.2,10) -- (axis cs:78.2,10);
\draw[cor,thick] ([yshift=-3pt]axis cs:62.2,10) -- ([yshift=3pt]axis cs:62.2,10);
\draw[cor,thick] ([yshift=-3pt]axis cs:78.2,10) -- ([yshift=3pt]axis cs:78.2,10);
\node[rectangle,fill=cor,minimum size=6pt,inner sep=0pt] at (axis cs:70.8,10) {};
\node[anchor=west,font=\scriptsize\bfseries] at ([xshift=4pt]axis cs:78.2,10) {70.8\%};
\end{axis}
\end{tikzpicture}
\caption{At equal size (\S{}12.4): share of champions among up to twenty topics a year, 2017--2022, taken for each funder as the candidate topics named in the most grants, with 95\% Wilson intervals. The dashed line is chance weighted like the list (22.3\%); the square is the detector's list, 70.8\% [62.2--78.2]. CZI names fewer than twenty candidate topics a year, so its line holds all the topics it named; Gates and CZI are measured on titles only. With groups of this size, differences below about ten points are within noise.}
\label{fig:4}
\end{figure}

\needlines{17}
\subsect{12.5\texorpdfstring{\quad}{ } Within the portfolio}
Among the candidate topics each funder was already funding:

\nopagebreak[4]
\begin{center}\small
\begin{tabular}{@{}lr@{}}
\toprule
funder & champions among the funded topics the list also designates \\
\midrule
Wellcome & \textbf{78.3\%} \\
NIH, record by record & 70.6\% \\
NIH, by project (first variant) & 72.5\% \\
UKRI & 75.8\% \\
European Commission & 76.7\% \\
ANR & 77.8\% \\
NSF & 73.8\% \\
Gates Foundation (titles only) & 63.2\% \\
CZI (titles only) & 81.8\% \\
\bottomrule
\end{tabular}
\end{center}

The split is the same for all eight funders: funded topics that the list does not designate fall back to the population's rate, or below it. For Gates and for NIH dated record by record, the topics the list designated but the funder did not fund did at least as well as those both designated and funded. Two cautions. \textbf{The list is narrow}: many of the champions present in each funder's portfolio are not on it. \textbf{Part of the gap is mechanical}: removing the best topics from a group lowers the rate of what remains.

\needlines{17}
\subsect{12.6\texorpdfstring{\quad}{ } Money arrives early}
For the 63 distinct champion names of the list, we compared the first grant naming each topic with the date of its first slot. The median lead is taken only over the champions that the funder named before the list.

\nopagebreak[4]
\begin{center}\small
\begin{tabularx}{\textwidth}{@{}P{3.6cm}R{2.2cm}>{\raggedleft\arraybackslash}XR{3cm}@{}}
\toprule
funder & first grant before the list & median lead among those & never named, under any form checked \\
\midrule
Wellcome & 23 of 63 & 43 months (38 without two short acronyms) & 11 \\
NIH, record by record & 56 & 57.6 months (52.6 without two short acronyms) & 0 \\
NIH, by project (first variant) & 58 & 87.6 months, inflated by construction & 0 \\
UKRI & 48 & 48 months & 1 \\
European Commission & 46 & 29 months & 1 \\
ANR & 32 & 31 months & 3 \\
NSF & 42 & 47 months & 3 \\
Gates Foundation (titles only) & 8 & 26 months (20.5 without one short acronym) & 34 \\
CZI (titles only, 2018--2024) & 4 & 24 months & 41 \\
\bottomrule
\end{tabularx}
\end{center}

\textbf{When the money arrives first, it arrives two to five years before the list.} Short acronyms (four characters or fewer) can match unrelated names, so the medians are also given without the short acronyms among the list's champions.

\textbf{The NIH dating, corrected.} Our first NIH variant dated a project by its first year but read it with the text of its later renewals: a 2010 project renewed in 2020 carried a 2020 abstract dated 2010. The third addendum added a variant in which each annual record is dated by its own budget start and read with its own text only. \textbf{The median lead falls from 87.6 to 57.6 months}: the first figure was about thirty months too long. NIH remains the earliest funder, both by median lead and by the number of listed champions it named before the list (56 of 63). Its selectivity barely moves (\S{}12.3).

\textbf{Never named.} The last column counts only verified cases: a champion counts as never named by a funder only if no grant of that funder carries any of the forms in our table, from 2010 to 2025. Some forms are broad, so this is the strictest reading. The three NSF cases are clinical topics, outside its mission. Most of the Gates and CZI cases lie outside their scope (oncology, computing, materials); with one sentence per grant, and for CZI seven years of data, we do not read them as blind spots.

\textbf{Does the list come before the money? No funder supports it.} A first count found ten or more listed champions that Wellcome, UKRI, the European Commission or ANR first funded only after the list. A check of former names and other forms in grant texts, applied to every funder, removed most of them. Wellcome funded \emph{Clostridium difficile} before the bacterium was renamed \emph{Clostridioides difficile}; NIH funded Enterobacteriaceae before Enterobacterales. The clean cases that remain are 6 for Wellcome, 3 for UKRI, 3 for the Commission, 7 for ANR, 1 for NIH, 4 for NSF, 4 for Gates and 6 for CZI; at CZI, none is a true lead, since four fall before its published grants begin and two in the year of the designation itself. None reaches the pre-registered threshold of ten. \textbf{No lead of the list over the money is claimed, for any funder} (Annex A, A20). Matching also errs in the other direction: three false matches were found (two short acronyms, one place name) and removed from the counts.

\needlines{11}
\subsect{12.7\texorpdfstring{\quad}{ } Is very early money more often wrong?}
This question was added during the study and measured under a second addendum, fingerprinted before computing. For each funder, funded candidate-years were grouped by the age of the first grant naming the topic at the photograph. The rule, applied to Wellcome only, required three conditions for the earliest group: a champion rate at or below the population's (20.2\%), at least 10 points below the latest group, and at least 30 cases in each group.

\nopagebreak[4]
\begin{center}\footnotesize\setlength{\tabcolsep}{5pt}
\begin{tabularx}{\textwidth}{@{}Yrrrrrrrrr@{}}
\toprule
first grant naming the topic, before the photograph & Wellcome & NIH, record & NIH, project & UKRI & EC & ANR & NSF & Gates & CZI \\
\midrule
five years or more & \textbf{18.2\%} & 17.4\% & 21.6\% & 23.3\% & 16.1\% & 19.2\% & 21.2\% & 16.3\% & --- \\
two to five years & 28.6\% & 26.7\% & 22.3\% & 25.8\% & 34.5\% & 29.1\% & 32.4\% & 23.3\% & 33.3\% \\
less than two years & \textbf{31.9\%} & 26.6\% & 18.7\% & 26.3\% & 27.6\% & 30.1\% & 25.6\% & 30.6\% & 35.3\% \\
\bottomrule
\end{tabularx}
\end{center}

\emph{(champion rates; every group of the five funders of the first study and of NSF holds several dozen candidate-years or more; the groups of Gates and CZI are much smaller, and CZI's data cannot hold a grant five years before a photograph; population 20.2\%)}

\textbf{For Wellcome, all three conditions hold.} The intervals of the earliest and latest groups overlap: the rule is met, but the gap is not far from noise. The latest group's topics are larger in recent volume than the earliest group's, but by less than a factor of two, so size does not explain the gap. The effect also holds without single-word names, which are the most ambiguous (19.8\% against 31.4\%); this check was made after the fact. The other funders are reported without a sentence, as the rule required. Dated record by record, NIH shows a gap in the same direction, which its first variant did not show; we report it without a conclusion.

Two reservations. Five years ahead, a topic's name is often not settled, and an early grant may use it in another sense; we have not measured how much of the effect this explains. And \textbf{a second pre-written measure does not show the same gradient.} Dating the first grant against the year the topic first appears in the literature, rather than against the photograph, Wellcome gives 21.4\% champions when money came before the topic appeared, 33.3\% in its first two years, and 22.2\% later.

\subsect{12.8\texorpdfstring{\quad}{ } What this study shows, and what it does not}
\textbf{What the measurement says, and only that.} Under the exact name matching of \S{}12.2, the detector's lists are more concentrated in future champions than the twenty topics named in the most grants of each funder: 70.8\% [62.2--78.2] against 20.8\% [14.5--28.9] to 35.8\% [27.8--44.7], with chance at 22.3\%. The interval of the detector overlaps none of the funders'.

\textbf{This result depends on exact matching by names.} A grant often describes a topic in its own words, through concepts that do not carry the topic's name, and exact matching misses such grants; it may therefore disadvantage grant texts against a list made of topic names. The comparison is a measurement within the protocol, not evidence that the detector selects better than funders. A funder, moreover, does not try to maximise the growth of a literature; the judge is ours, the candidate pool is ours, and the matching has not been audited by hand. Within the noise, the rest is description: grant money from eight funders, public and private, arrives early and touches many candidates; within a portfolio, the funded topics that the list also designates hold far more often than those it does not; the one funder above the pre-registered bar owes it to one thematic bet, on a small sample; and once former names are checked, no funder has enough cases to say that the list comes before the money. Texts differ between funders in length and coverage, and two funders publish one sentence per grant.

\sect{13\texorpdfstring{\quad}{ } Limitations}
\textbf{The specification is not published} (\S{}7.2), and neither is the candidate population, which is part of the detector and carries much of the result. The judge's rule and the bets' fingerprints can be checked without an agreement; the benchmark slots and their verdicts, under agreement.

\textbf{The benchmark is retrospective} (Part B). Claims about future performance rest on bets 5 and 6, the only ones that test exactly its instrument; bet 5 is judged in early 2029, bet 6 in early 2030.

\textbf{The temporal control of the historical benchmark is imperfect.} Its photographs read a snapshot of 2026, some inputs of the program were prepared after the photograph dates, and the photograph overlaps the judge's denominator (Part B). We measured the size of revisions for one recent period only (\S{}4.1).

\textbf{For future bets, what we plan.} We plan to accompany each future bet with a \textbf{complete, dated snapshot of the data at the date of the photograph}, fingerprinted, and to have it \textbf{executed, or its sealed materials held, by a third party}.

\textbf{The bets were sealed late.} Bets 1 to 5 were sealed fourteen to fifteen months after their photograph, when the 2025 literature was already written. The programs read nothing after the photograph, but the snapshot itself is of 2026. Bet 6 reduces the delay to three months.

\textbf{One corpus.} All results are on PubMed/MEDLINE, and they depend on the material the detector reads (\S{}6.5). Outside biomedicine we have illustrations, not a score.

\textbf{The judge measures volume.} A champion may be a fashion. The rule can also be satisfied by generic topics or by renamings. Two of the topics in our benchmark examples changed name during the period.

\textbf{Epidemic vintages.} The 2021 and 2022 vintages are dominated by the COVID-19 literature. Failures are four times more frequent among tracked topics in 2022 than in 2017. All three failures of the detector fall in these vintages.

\textbf{Stability of the lists was never measured.} Small changes in the data can reorder a list: bet 6 found eight of twenty positions reordered after the updates. Two implementations of the same specification give the same totals with seventeen different names among the champions. \textbf{Totals are robust from one implementation to another; names are not.}

\textbf{Some champions are out of reach by construction} (about thirty-five in an earlier state of the benchmark). Their re-measurement is still due.

\textbf{The judge is ours.} Nine variants of it, each changing one parameter, keep the advantage (\S{}10); the longest horizon is the weakest (2.5 times chance). This robustness is shown on the retrospective benchmark only.

\textbf{No comparison with modern semantic methods.} We did not compare the detector with methods built on semantic representations of texts (embeddings), such as those of [34] or the recent benchmarks [27, 28]. These are the methods of the current state of the art. We claim nothing about how the detector would fare against them.

\textbf{No documented comparison with an independent effort.} Apart from chance, the public methods of \S{}5.1 (run on all tracked topics) and the funder lists of \S{}12, this paper reports no score for another detector on the same benchmark. Public methods run inside the candidate population (\S{}5.3) are reported without counts (\S{}7.2). Whether a serious independent effort, given the rules and a held-out period, would match the detector has not been measured in a form we can document.

\textbf{The anchoring is public; the manifests are not.} A reader can check that each manifest existed at its date, but can check that a given fingerprint is in it only under agreement (\S{}11.4).

\textbf{The funder study is secondary and exploratory} (\S{}12.1). Its pre-registrations were not externally timestamped, and its exact name matching, not audited by hand, may disadvantage grant texts (\S{}12.8). Two funders publish one sentence per grant, and one of them only from 2018.

\textbf{Our own interest.} The author has a financial interest in Loxyn SAS, which holds the specification and sells lists produced by the detector. The sealed bets were designed so that this interest cannot rewrite a list or its verdict rule after the fact; it does not remove the fact that the candidate pool, the judge and the retrospective benchmark were designed by an interested party.

\sect{14\texorpdfstring{\quad}{ } Conclusion}
The main contribution of this paper is a protocol: before any verdict can be known, freeze the date of the photograph, the population of candidates, the verdict rule, and the comparison with chance among the same candidates; then seal. The protocol does not depend on our detector. It applies to any tool that publishes lists of emerging topics, and it stays defensible if our bets fail: a failed bet, judged by a rule written in advance, is exactly what the protocol is built to show. Sealing proves that a list existed at a date, not that it is right.

Applied to one detector, the protocol keeps three things apart. \textbf{Exploration} shows how many paths were tried and that most failed (Part A). \textbf{Historical validation} gives 85 champions and 3 failures out of 120 slots, against 26.7 and 15.1 for chance among the same candidates, and the advantage holds under nine variants of the judge (Part B); this is retrospective evidence, developed with the verdicts known. \textbf{Prospective validation} is still to come: the bets will be judged in early 2029 and early 2030, and any claim about future performance waits on bets 5 and 6, the only ones that test exactly the instrument of the benchmark (Part C). The secondary study of research funders says one thing, under exact name matching: the detector's lists are more concentrated in future champions than the topics named most in grants (\S{}12). That matching may disadvantage grant texts, which often describe a topic without naming it; the result is a measurement within the protocol, not evidence of better selection.

We plan to publish an open toolkit that implements the protocol independently of our detector (Data and code availability).

\textbf{We invite replication} under agreement, on PubMed/MEDLINE (Data and code availability).

\sect{Data and code availability}
Public now: the judge's rule (\S{}3.2) and its nine variants (\S{}10), the fingerprints of the sealed objects (Annex B), and the funder data sources (\S{}12.2). Available under agreement: the 120 judged slots of the benchmark with their verdicts, checkable against their fingerprint. Not published: the specification of the detector, including the candidate population (\S{}7.2). OpenTimestamps proofs of the anchored manifests and public files of position fingerprints: doi:\url{10.5281/zenodo.23641352} (Zenodo); the manifests themselves: under agreement (\S{}11.4).

\textbf{Code.} We plan to publish an open toolkit that implements the protocol independently of our detector: the judge, computed from a topic-by-year table supplied by the user; chance and the ceiling among candidates supplied by the user; the sealing; and an open reference detector, burst detection [15], applied to a public and neutral layer such as the topics of OpenAlex [38], with no data and no component of our detector.

\textbf{Replication.} Any independent group is invited to an evaluation under agreement on PubMed/MEDLINE: the group chooses the sub-domain, the period and the measure, receives sealed lists, and judges them itself.

\sect{Use of AI tools}
AI agents were used throughout this work: to run measurements, to perform the reverse-engineering tests (\S{}7.2), to review our documents adversarially, and to draft this manuscript. Numbers were checked against the output file or the dated verdict that produced them; an internal table maps each number to its source. A few figures rest on a dated verdict without a separate data file; they are qualitative in the text.

\sect{Competing interests}
The author has a financial interest in Loxyn SAS (see \S{}13).

\sect{Annex A\texorpdfstring{\quad}{ } Register of corrections}
We keep a register of every departure from the letter of our procedural rules, however benign, and of every correction that changed a published or circulated figure. Each entry gives its date, its scope and its repair. Nothing is removed from the register. Failing being able to show everything, we show that we note everything.

Entries are filed under the three blocks of the paper (\S{}1.4). Entries about the funder study, which is exploratory (\S{}12), are filed under exploration. Within each block, entries keep their number and their order.

\subsect{A.1\texorpdfstring{\quad}{ } Exploration}
\textbf{A1 \textperiodcentered{} 20-08-2026 \textperiodcentered{} the judge's data seen before a sealed design.} The day before a component was designed and sealed, a check read volumes of one topic in the data of its future verdict (years and counts, no statistic of the judgement). The component later failed its trial, and the clause ``no reinterpretation'' was applied. Declared in the sealed design and in its verdict.

\textbf{A5 \textperiodcentered{} 31-08-2026 \textperiodcentered{} a component half wired.} The first measurement of a candidate component applied only half of its definition. The result, wrongly unfavourable by ten champions, nearly counted against it. Repair: complete definition rewired, both versions recorded.

\textbf{A12 \textperiodcentered{} 11-09-2026 \textperiodcentered{} country detection by substring.} Our first country detection in affiliations searched for substrings (31.5\% of records marked British were false; \S{}4.2). Country counts computed before 11-09-2026 are not reliable for the United Kingdom. This affects the country counts printed for two named topics in our earlier public evidence file (8 September 2026), which this paper supersedes on this point; they are not used in this paper.

\textbf{A14 \textperiodcentered{} 03-10-2026 \textperiodcentered{} funder study, four corrections.} A coding defect assigned some grant matches to only one of the topic names they matched; corrected after a first run, no conclusion changed. A first timing figure was a signed median over topics funded before and after the list, which does not measure a lead; it was replaced by the median among topics funded before the list (38 to 43 months). A first count of twenty topics where the list came before Wellcome did not survive a check of former names; four to seven remain. The equal-size comparison was added after an internal review.

\textbf{A18 \textperiodcentered{} (August 2026) \textperiodcentered{} the sofosbuvir case re-anchored.} A first version of the founding case gave September 2011 on a seed of six articles. Its scripts, at first believed lost, were found: they measured the whole family of molecules with the suffix ``-buvir'', with a different reference. It was not the same object. The case was re-anchored on the construction whose every number replays to the byte (July 2011). Rule: a figure that does not replay does not open an article, even if it is true.

\textbf{A20 \textperiodcentered{} 08-10-2026 \textperiodcentered{} the lead of the list falls for every funder (second erratum to the funder verdict).} The funder verdict of 3 October said that the list came before the money for Wellcome, UKRI, the European Commission and ANR: ten or more listed champions first funded only after the list. A first erratum, the same day, removed Wellcome (A14). On 8 October, the check of former names and other forms in grant texts was applied to every funder. It reduced the clean cases to 3 for UKRI (from 11), 3 for the Commission (from 11), 7 for ANR (from 22), 6 for Wellcome and 1 for NIH. No funder reaches the threshold of ten: no lead of the list is claimed for any funder. The other conclusions (the list chooses better than the money, at every size) hold for all nine variants. Declared as an erratum appended to the verdict, which was not rewritten.

\textbf{A21 \textperiodcentered{} 08-10-2026 \textperiodcentered{} the dating of NIH grants.} Our first NIH variant dated a project by its first year but read the text of its later renewals. A variant dated record by record, added by the third addendum, gives a median lead of 57.6 months instead of 87.6. Both are reported (\S{}12.6).

\textbf{A22 \textperiodcentered{} 08-10-2026 \textperiodcentered{} two false matches found after the fact.} A short acronym matched an unrelated organisation in a Gates grant, and was one of its clean cases; it is removed from that count. A topic name matched a grant to a county in the CZI list, and was one of its champions; the rule's figure keeps it, and the rate without it is given (\S{}12.3).

\textbf{A23 \textperiodcentered{} 08-10-2026 \textperiodcentered{} two scripts rewritten.} The scripts of the first two addenda were not kept. They were rewritten from the sealed text and reproduce the published results of the first five funders exactly. A simpler computation of chance weighted like the list in the rewritten script, twenty times each vintage's rate, gives 27.0 champions and 15.0 failures instead of 26.7 and 15.1. No rule depends on it.

\subsect{A.2\texorpdfstring{\quad}{ } Historical validation (the 2017--2022 benchmark)}
\textbf{A4 \textperiodcentered{} 30/31-08-2026 \textperiodcentered{} one script, one output.} Four benchmark scripts wrote to the same output file, so each run overwrote the evidence of the previous one. No fingerprint was lost. Repair: outputs renamed, full re-run, fingerprints identical to the byte.

\textbf{A7 \textperiodcentered{} 01-09-2026 \textperiodcentered{} a validation that read the future.} The validation of a revision of the instrument read the whole series to 2025 for every vintage (\S{}6.4). Measured the same day without the future; the part responsible was replaced. Bet 2, already sealed with it, runs as sealed.

\textbf{A9 \textperiodcentered{} 01-09-2026 \textperiodcentered{} slots counted as objects.} Several figures of the benchmark had been counted in slots while being described as distinct objects. Nine per cent of slots were redundant. An earlier figure of 88 champion slots covered 80 distinct objects; earlier figures of 82 and 87 were affected in the same way. Repair: the counting rule for equivalent names was applied to the benchmark (85/3, sealed 04-09-2026), and figures are now given in slots with the number of distinct names (63; 59 after merging equivalent names).

\textbf{A10 \textperiodcentered{} 07-09-2026 \textperiodcentered{} names compared with slots.} An internal note compared a number of distinct names from one benchmark with a number of slots from another. Rule since then: every benchmark figure is given in names and in slots, never one against the other.

\textbf{A11 \textperiodcentered{} 08-09-2026 \textperiodcentered{} names are fragile.} Two implementations of the same specification gave the same totals (85/3) with seventeen different names among the champions. Rule: totals are robust across implementations, names are not. Our earlier public evidence file cites 63 distinct names, the count of the sealed file.

\textbf{A13 \textperiodcentered{} 12-09-2026 \textperiodcentered{} two conventions in our own tables.} The defect described in \S{}4.2 (703 entries) came from our own tables, not from the source. Measured without effect on the world lists; not measured on the retrospective benchmark.

\textbf{A16 \textperiodcentered{} 07-10-2026 \textperiodcentered{} which version of a component gives 85.} Internal texts attributed the sealed 85/3 to the later revision of a component of the specification. A recomputation showed that the sealed file uses the earlier version, the same as the six bets; the later revision gives the same 85/3 with two slots changed (\S{}8.2).

\textbf{A19 \textperiodcentered{} 07-10-2026 \textperiodcentered{} two wordings in our public evidence file.} Our earlier public evidence file (8 September 2026), which this paper supersedes on these points, describes the 85 champion slots as ``all distinct objects''; they carry 63 distinct names, 59 after merging equivalent names (\S{}3.6). The same file names one position of the dated register ``cardiovascular-renal-metabolic syndrome''; the official name, used here, is cardiovascular-kidney-metabolic syndrome.

\textbf{A26 \textperiodcentered{} 24-10-2026 \textperiodcentered{} counts at sixty places withheld.} The pre-registration of \S{}9 stated that the paper would print chance, the ceiling and the result at twenty and at sixty places. At sixty places these counts are withheld with the specification (\S{}7.2). They remain in the fingerprinted results file (Annex B). The comparison itself was run as pre-registered; no figure changed.

\textbf{A24 \textperiodcentered{} 09-10-2026 \textperiodcentered{} a control written with the wrong rounding.} The pre-registered control of the sensitivity measurement (\S{}10) required the reference judge to give back 26.7/15.1. The program first compared a value rounded to one decimal, which floating-point arithmetic rounded the wrong way, and stopped. The control was rewritten to compare the stored, unrounded values, of which 26.7/15.1 is the printed form. A first stop had come from a programming error on the 2022 vintage under variant J6, outside its declared scope. No figure changed between runs.

\subsect{A.3\texorpdfstring{\quad}{ } Prospective validation (sealing and bets)}
\textbf{A2 \textperiodcentered{} 25-08-2026 \textperiodcentered{} an anchor that no longer matched the disk.} A document was rebuilt after it had been fingerprinted and anchored the same day; the anchored fingerprint matched no file on disk. Nothing false was anchored. Repair: a new manifest on 26-08, earlier ones left intact, the gap written into the new manifest.

\textbf{A3 \textperiodcentered{} 25-08-2026 \textperiodcentered{} a sealed file corrected on its day of deposit.} The revelation register of datings was corrected on the day it was deposited. Its first fingerprint (\texttt{85a3e459\dots{}}) became void and was replaced by \texttt{8f3b7815\dots{}}. Repair: a separate erratum the same day, the old fingerprint declared void in full.

\textbf{A6 \textperiodcentered{} 31-08-2026 \textperiodcentered{} a column header that invited a wrong reading.} The register of datings had a column headed in a way that suggested the sealing dated from the dating. An email repeated the misreading (``a position sealed seven years before''). The sealing dates from 25-08-2026; the seven-year lead is that of a back-computed dating. Repair: column renamed, erratum of reading, reference wording fixed for every public version.

\textbf{A8 \textperiodcentered{} 01-09-2026 \textperiodcentered{} lists that repeated a topic.} The four lists of vintage 2025, first sealed on 31-08 and 01-09, held the same topic in more than one position. Decision: the four bets were withdrawn and resealed on 01-09 under a declared rule. This replacement is legitimate for one reason only: none of the lists had been communicated to anyone, and the verdict is in 2028. The earlier manifests remain intact and anchored; each removed position is named in its deposit. Final state: manifest v19.

\textbf{A15 \textperiodcentered{} 07-10-2026 \textperiodcentered{} retired fingerprints in an internal draft.} An internal draft of this article printed fingerprints of bets 1 and 2 that belonged to withdrawn versions (A8). The correct fingerprints are those of Annex B.

\textbf{A17 \textperiodcentered{} 07-10-2026 \textperiodcentered{} the non-regression of bet 6 failed, and was declared.} With the update files, the frozen program did not give back bet 5 exactly (same topics, eight positions reordered; one internal control changed by one champion). The computation was stopped, the choice was referred to the author, and bet 6 was pre-registered with this amendment declared before computing (\S{}11.2).

\textbf{A25 \textperiodcentered{} 24-10-2026 \textperiodcentered{} the verdict date read on complete volumes.} The pre-registrations of bets 1 to 5 set the verdict at 31 December 2028. Since a month needs about three months to be indexed (\S{}4.1), the 2028 volumes will be read on the first PubMed baseline that contains the complete year 2028 (early 2029), and likewise for bet 6 and the year 2029. The verdict rule itself is unchanged. This reading rule was already printed in our text of 7 October 2026; it is recorded here on 24 October 2026. No verdict can be known before 2029.

\sect{Annex B\texorpdfstring{\quad}{ } Fingerprint table}
All fingerprints are SHA-256. Where only a prefix is given, the full value is in an anchored manifest, shown under agreement (\S{}11.4).

\needlines{9}
\textbf{Substrate and benchmark}

\nopagebreak[4]
\begin{center}\small
\begin{tabularx}{\textwidth}{@{}YP{9.9cm}@{}}
\toprule
object & fingerprint \\
\midrule
manifest of the 1,334 PubMed baseline files (frozen 18-08-2026) & \texttt{615912ccc2ee341c\dots{}} \\
benchmark output, 120 judged slots, 85/3 (04-09-2026; recomputed identical 07-10-2026) & \fp{d059c3f0cd093b81c8e3daf6e121a92d3090feafea2426632918e27479951258} \\
sealed file of the rarity of the judge & \fp{617823f3ed8c2dd8b81aacce4d0d64846bc1fb2e04208788a8a60fc0a0f6f037} \\
revelation register of datings (25-08-2026) & \fp{8f3b78151f802aa899907661fadbdf65c0586a45eb33f4ae63807f5a1b4f18f7} \\
\bottomrule
\end{tabularx}
\end{center}

\needlines{10}
\textbf{Bets 1 to 4} (vintage 2025, photograph 30-06-2025, sealed 01-09-2026)

\nopagebreak[4]
\begin{center}\small
\begin{tabularx}{\textwidth}{@{}YP{9.9cm}@{}}
\toprule
bet & list fingerprint \\
\midrule
1 \textperiodcentered{} frozen protocol & \fp{745f5946d3fd955ff17c09acd5901a2850e391a7b5927694dec75cdfd891e286} \\
2 \textperiodcentered{} revised instrument & \fp{9dfd3c80ba0d65f00f87661a4da17c5fe488b34d52061f2304f5eb75f9182b46} \\
3 \textperiodcentered{} third revision & \fp{b8c731b34fca431a03ed4988b1b98f36e95af7a758e58d2a21a59b5f66e009bb} \\
4 \textperiodcentered{} the full score & \fp{a430880f2e897ecb7ba312029a9ff8b6a5d6e14e14d0dabe5bdff93026eb6dd9} \\
anchored manifest v19 & \fp{2ef27c4fe7bc731f4caa432a0da8b56f11a279b09e06c2fc47d77aafe3b6cd98} \\
\bottomrule
\end{tabularx}
\end{center}

\needlines{14}
\textbf{Bet 5} (vintage 2025, photograph 30-06-2025, sealed 07-10-2026)

\nopagebreak[4]
\begin{center}\small
\begin{tabularx}{\textwidth}{@{}YP{9.9cm}@{}}
\toprule
object & fingerprint \\
\midrule
pre-registration & \fp{c4cd25e6baf98fd2ff95de9158cd95bd7d8eae76d7eb0005f96b31665509307b} \\
frozen program (six files; fingerprint of the list of their six SHA-256, in execution order, one per line) & \fp{90d7d25c453fa31c8bfdb73d22aa768f8495b0d98a32bea37b8c90f4a3e6befe} \\
amendment 1 & \fp{145a8f664f5562e8f2771e0ffb807796f86e0877e842b9b6062e4dba6b55c4e8} \\
amendment 2 & \fp{a3067e7bec2f1b598f7573520612af2b4851c0f06bab04285387c25c23b24839} \\
bet file & \fp{49a5b9a50632769896a41bf955c68aedef31dabd35da99d769dc664bd9fad81d} \\
candidate population & \fp{6ed32f664d0b84fa666e52c8a4d12a877ef9862abaed6deb4b328bf00585ef17} \\
public file of the twenty position fingerprints & \fp{0dc919b0480041f59030fbe7b924eab54e5c3ab9ae0a152e40a64be02cf64a00} \\
root of the positions & \fp{2276d08214975d72c73a7281ff45656b836074713d32a4aa1b257ddb2ef98ab9} \\
anchored manifest v53 & \fp{24e916899b25fe98cd61844d8869fd2dcbac4bb824a5ef61eb549e51ac56679e} \\
\bottomrule
\end{tabularx}
\end{center}

\needlines{12}
\textbf{Bet 6} (vintage 2026, photograph 30-06-2026, sealed 07-10-2026)

\nopagebreak[4]
\begin{center}\small
\begin{tabularx}{\textwidth}{@{}YP{9.9cm}@{}}
\toprule
object & fingerprint \\
\midrule
pre-registration (amendment included) & \fp{74246d07c3b77a2e429dd66fcb12bb2552836fea56c3b53f1774d6325deaec91} \\
frozen program (six files; fingerprint of the list of their six SHA-256, in execution order, one per line) & \fp{a870afc0ef1ecad97d0a92f9edb112eb955254e92d3cf0700a35205ca0796c35} \\
bet file & \fp{310f1f1486eda05c6795d10dc8b333308ff5f55ca3eb4238e8a2d5082de25d28} \\
candidate population & \fp{1f1c1568f030ae7fc42f1261bac2645506b589d68c2f7a35d08aad64ff3095c6} \\
public file of the twenty position fingerprints & \fp{fc26b066e872a1d873fffade4ba83c0f6e55ab571833979f18127d99d0ab3de6} \\
root of the positions & \fp{c89e0d94fcca1c525746e7e62e4705c14b891c23c3c6f15c67079b8069dc60d9} \\
anchored manifest v54 & \fp{aac0131e07895d176099b8a69a758d98aabcbd82f7cedbfae801ec92acf17f46} \\
\bottomrule
\end{tabularx}
\end{center}

\needlines{8}
\textbf{Twenty against sixty places, \S{}9} (07-10-2026; fingerprinted, not externally timestamped)

\nopagebreak[4]
\begin{center}\small
\begin{tabularx}{\textwidth}{@{}YP{9.9cm}@{}}
\toprule
object & fingerprint \\
\midrule
pre-registration (written before computing) & \fp{634de397b9a419e6d29e2e677ff7f885bd36f1929b153d857d6443c11741967e} \\
list output, sixty places & \fp{e59d36ee3754382e971f2f37ab08faf52651616485e0a1ad417e602ee9583a97} \\
results, sixty places & \fp{85f0558cb1d869095482994fc93959c9af113f717de690e20818d8d2ee254a48} \\
\bottomrule
\end{tabularx}
\end{center}

\needlines{8}
\textbf{Sensitivity of the judge, \S{}10} (09-10-2026; fingerprinted, not externally timestamped)

\nopagebreak[4]
\begin{center}\small
\begin{tabularx}{\textwidth}{@{}YP{9.9cm}@{}}
\toprule
object & fingerprint \\
\midrule
pre-registration (written before computing) & \fp{5450b7ce98b743423dd952ab31b1bf1b3cb34b739d2cbd4837f37f6b0c0585c3} \\
results, ten variants & \fp{21d56e1513dd3134ebf127d21f55a0c5b3546ceebe9018aced6fec462c85010a} \\
list re-run for the control (identical to the sealed benchmark output) & \fp{d059c3f0cd093b81c8e3daf6e121a92d3090feafea2426632918e27479951258} \\
\bottomrule
\end{tabularx}
\end{center}

\needlines{11}
\textbf{Funder study} (03-10-2026 and 08-10-2026; fingerprinted, not externally timestamped)

\nopagebreak[4]
\begin{center}\small
\begin{tabularx}{\textwidth}{@{}YP{9.9cm}@{}}
\toprule
object & fingerprint \\
\midrule
main protocol & \fp{410bedca39dbf545805464e734de3b77dfce703e97e5f39867a8932b7b812baf} \\
addendum 1, equal size & \fp{58cd5ef5285a0790d4d6b66d8efa1ec9bec4b338c20be67b098c51df31dd5b5e} \\
addendum 2, very early money & \fp{c55c9bfdfb342e2ff4379a9ebe952cf75499cc07693c81de375227f80374cdad} \\
results file, first five funders & \fp{54819d359f61c21da77d7f442e60ded3257332cbd6389a6ea69dbb7f4134c202} \\
addendum 3, US funders and NIH record by record (08-10-2026) & \fp{56848256f3a873676a32be06be19f068831ed73d078f19f9e8de05090951cfb9} \\
results file, US funders (08-10-2026) & \fp{dca230d40f8576dd80f7e19c3d6fc33d23d1bfd932a9906f409a687a44956e8c} \\
\bottomrule
\end{tabularx}
\end{center}

\sect{Annex C\texorpdfstring{\quad}{ } The funder protocol in brief}
\begin{enumerate}
\item \textbf{Photograph} \emph{T} = 30 June of vintage \emph{Y}, 2017--2022. ``Before \emph{T}'' means strictly before. A grant counts only if dated from 2010-01-01 to 2025-12-31.
\item \textbf{Candidates} are the candidate-years of the benchmark; their verdicts are those of \S{}3.2.
\item \textbf{Matching}: whole expression in title or description, lower-cased, punctuation removed; the last word may take a final ``s''; no synonyms, no language model.
\item \textbf{Measure A, selectivity}: champion and failure rates among candidates funded before \emph{T}; main variant (any grant since 2010), and two secondary variants (grants of the last 24 months; at least three grants).
\item \textbf{Measure B, timing}: for each distinct champion of the list, first grant naming it against the date of its first slot; the median lead is taken over the champions funded before the list only.
\item \textbf{Measure E, combination}: candidates funded and listed, listed not funded, funded not listed.
\item \textbf{Reading rules} (written before): money ``chooses better than the population'' at $\geq$ 1.5 times its champion rate; the list ``chooses better than the money'' at $\geq$ 20 points more champions, with no more failures in proportion; a lead of the list over a funder is claimed only if at least ten listed champions were first named by that funder after the list; only the main variant decides; everything else is reported without a sentence.
\item \textbf{Addendum 1 (equal size)}: each funder's up to twenty candidates per year with the most grants before \emph{T} (ties broken by most recent grant, then name); same rule of 20 points.
\item \textbf{Addendum 2 (very early money)}: funded candidate-years grouped by age of first grant (\textless{} 24 months, 24--60, $\geq$ 60) and, separately, by first grant against the topic's first year in the literature; the rule of \S{}12.7, applied to Wellcome only; size control by median recent volume.
\item \textbf{Addendum 3 (US funders, 8 October 2026)}: NSF, the Gates Foundation, the Chan Zuckerberg Initiative, the Burroughs Wellcome Fund and HHMI, under the same protocol; admission at $\geq$ 1,000 grants dated 2010--2025 with a text, at least half of them with $\geq$ 50 characters; funders with one sentence per grant measured and marked ``titles only''; a variant of NIH dated record by record, reported beside the first; a table of the other forms a grant text may use for each listed champion (former names, long forms of acronyms, British forms, drug code names), written before the US results were read and applied to every funder: a lead of the list counts only if no grant of the funder carries one of these forms before the designation, and a champion counts as never funded only if none carries one from 2010 to 2025.
\end{enumerate}

\end{document}